\documentclass{aastex631}

\begin{document}

\title{Ion Charge States from a Global Time-Dependent Wave-Turbulence-Driven Model of the Solar Wind: Comparison with in-situ Measurements}

\author[0000-0002-1859-456X]{Pete Riley}
\affiliation{Predictive Science Inc. \\
9990 Mesa Rim Road, Suite 170 \\
San Diego, CA 92121, USA}

\author[0000-0001-9231-045X]{Roberto Lionello}
\affiliation{Predictive Science Inc. \\
9990 Mesa Rim Road, Suite 170 \\
San Diego, CA 92121, USA}

\author[0000-0002-8748-2123]{Yeimy J. Rivera}
\affiliation{Center for Astrophysics  $\vert$ \\
Harvard \& Smithsonian \\
60 Garden Street \\
Cambridge, MA 02138, USA}



\begin{abstract}

Solar wind charge-state measurements contain a wealth of knowledge related to the properties of the solar corona from where they originated. However, their interpretation has remained challenging because it convolves coronal temperature, density, and velocity along the particles' trajectory through the corona before they ``freeze in'' and are convected outward through the solar wind. In this study, we calculate ion charge states by coupling a non-equilibrium ionization model with a global magnetohydrodynamic model of the corona and inner heliosphere. We present results for two periods characteristic of solar minimum and maximum and compare them with observations from the ACE spacecraft. We find that the model reproduces the essential features of the observations, rectifying an earlier inconsistency that was apparent in 1-D calculations, and allows us to unambiguously trace the evolution of charge states from the base of the corona into the solar wind. 

\end{abstract}



\section{Introduction}

The solar wind, a continuous flow of charged particles emanating from the Sun, plays a pivotal role in shaping heliophysical phenomena and the dynamics of the solar system (e.g., \cite{cranmer17a}). Originating from the solar corona, the plasma propagates through interplanetary space, carrying with it the solar magnetic field \citep{2013LRSP...10....5O}. Ultimately, understanding the dynamics of the solar wind is crucial for revealing the fundamental processes of solar-terrestrial interactions and for predicting space weather events that can impact satellite operations, communication systems, and ground-based technologies \citep{2006LRSP....3....2S}. As such, a comprehensive understanding of solar wind dynamics is indispensable for advancing our knowledge in heliophysics and protecting our technological infrastructure.

The charge states of ions within the solar wind offer critical diagnostics for understanding the properties and dynamics of the source regions in the solar corona \citep{Rivera2022} and have been used as markers that link heliospheric structures to their coronal sources \citep{Brooks2015, Parenti2021, Ervin2024, Rivera2024b}. These charge states, which are typically frozen into the plasma as it escapes the Sun's gravitational and magnetic fields, reflect the temperatures and densities at which they were formed \citep{hundhausen68a}. For instance, high charge state ratios such as O$^{7+}$/O$^{6+}$ are indicative of hotter source regions and have been used effectively to map the thermal structure of the corona during various solar activities \citep{Landi2014}. Furthermore, variability in the charge states can reveal changes in the coronal heating processes and the mechanisms driving the solar wind \citep{landi12a, Gilly2020}, as well as the complex evolution taking place when a coronal mass ejection (CME) erupts \citep{rakowski07a,gruesbeck12a,rivera19a,laming23a}.  

Previous studies employing Magnetohydrodynamic (MHD) modeling have significantly advanced our understanding of solar wind dynamics and the evolution of charge states in the solar atmosphere and heliosphere \citep{2015ApJ...806...55O,2022ApJ...926...35S,2023ApJ...959...77L,2023ApJ...955...65R}. Recent advancements have seen these models become increasingly sophisticated, incorporating detailed chemistry and complex magnetic field structure to predict the charge states in the solar wind. 

In this study, we present a loosely coupled 3-D charge-state/MHD model and develop global heliospheric solutions for two time periods of interest; one in the declining phase of the solar cycle (Carrington Rotation (CR) 2063) and one closer to solar maximum (CR 2002). For each solution, we explore the features of the charge-state distributions in various types of solar features and compare their profiles with ACE measurements at 1 AU in the ecliptic plane. In the next section, we summarise the data sets analyzed and the models used. In Section 3, we describe the results of the simulations and compare them with in-situ measurements of plasma, magnetic field, and charge states. In Section 4, we discuss the implications of these results, and finally, in Section 5, we make some general conclusions and suggest several avenues for potentially fruitful studies in the future. 

\section{Methodology}
\subsection{Observational Data}

In this study, we use data from NASA's Advanced Composition Explorer  \citep[ACE;][]{Stone1998} mission, and in particular, in-situ composition, plasma, and magnetic field measurements.
We use two-hour integrated measurements of heavy ion composition data from the Solar Wind Ion Composition Spectrometer (SWICS; \citealt{Gloeckler1992}). We use bulk plasma properties from Solar Wind Electron, Proton, and Alpha Monitor (SWEPAM; \citealt{Mccomas1998}), and magnetic field measurements from the magnetic field experiment (MAG; \citealt{Smith1998}) from the merged 64-second temporal resolution dataset\footnote{\url{https://izw1.caltech.edu/ACE/ASC/level2/lvl2DATA_MAG-SWEPAM.html}}. 

\begin{figure}[ht!]
    \centering
    \includegraphics[width=0.975\textwidth]{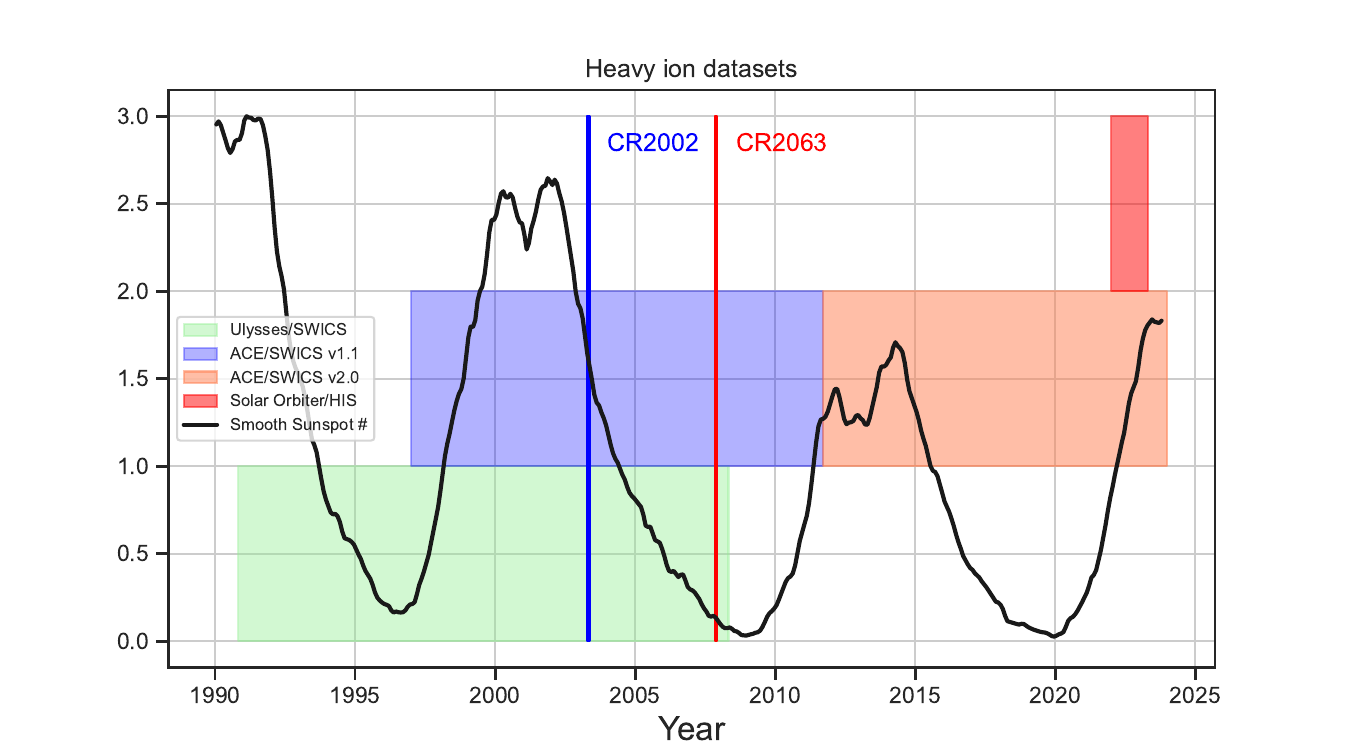}
    \caption{Timeline showing the availability of heavy ion in-situ measurements. The transition from ACE/SWICS v1.1 to V2.0 is indicated by the transition from purple to pink. The monthly averaged, smoothed sunspot number is overlaid with the solid black line. Carrington rotations 2002 and 2063 are marked by the vertical blue and red lines, respectively.  
}
    \label{data-availability}
\end{figure}

Figure~\ref{data-availability} summarises the main NASA missions that included a heavy ion detector. The Ulysses and ACE SWICS instruments provide the most overlap for cross-calibrated measurements since the instruments and software pipelines are similar. Since 2011, ACE/SWICS has continued to produce heavy ion data, but the data (v2.0) is limited in extreme conditions \citep{swics24a}. Previous work has analyzed periods with similar caveats that will apply to this study \citep{Rivera2020}. Most recently, the HIS instrument onboard Solar Orbiter provided data between January 2022 and April 2023. A comparison between SO/HIS and ACE/SWICS ion ratios and elemental composition found similar correlated behaviour across solar wind speed showing compatibility across datasets \citep{Livi2023}. 

In this study, we focus on two specific Carrington Rotations (CRs) occurring during ACE/SWICS's v1.1 interval: CR2002 and CR2063. CR2002 spans from April 15, 2003, to May 16, 2003, and CR2063 spans from November 4, 2007 to December 2, 2007. The former represents a more active period shortly after the peak of the solar cycle, and the latter represents the late declining phase, just before the minimum of cycle 23. 

\subsection{Model Description}

As described by \citet{2019SoPh..294...13L} for a 1-D implementation, we loosely coupled our MHD code with a charge-state evolution code. In this 3-D extension, we ran the global MHD model forward in time until a steady-state equilibrium had been reached. We then used the velocity, density, and temperature fields to evolve the fractional charge states from the solar surface to 1 AU (although the evolution occurs until they become frozen in at several $R_S$, and then the values are advected beyond this point). In the subsections below, we describe each element in more detail. 


\subsubsection{MHD Model}

To model the global structure of the solar corona and heliosphere, we use Predictive Science Inc.'s (PSI's) Magnetohydrodynamic Algorithm outside a Sphere (MAS). MAS is a 3-D thermodynamic model used for simulating coronal structure, dynamics, and coronal mass ejections \citep{1999PhPl....6.2217M,1999JGR...104.9809L,2009ApJ...690..902L,
2013Sci...340.1196D,2018NatAs...2..913M,2005ApJ...625..463L,2006ApJ...642L..69L,2011ApJ...731..110L,2003PhPl...10.1971L,2013ApJ...777...76L,
2018ApJ...856...75T}. The MAS code integrates the time-dependent resistive thermodynamic magnetohydrodynamic (MHD) equations in spherical coordinates. This model has been continuously developed and improved over 30 years, with significant advancements in its capabilities and applications.

One of the key features of the MAS model is its Wave-Turbulence-Driven (WTD) heating capabilities, which have been instrumental in improving the realism of coronal simulations \citep{2018NatAs...2..913M,2023ApJ...959...77L,2023ApJ...959L...4M}. As described by \citet{2018NatAs...2..913M}, the WTD model incorporates the effects of Alfv\'en wave turbulence in heating the corona and accelerating the solar wind. The model has been described in more detail in  Appendix A of \citet{2018ApJ...856...75T}. Here we focus on a few key points that illustrate its strengths and limitations.

First, the WTD approach allows for a more accurate representation of coronal temperatures and solar wind speeds, addressing long-standing challenges in coronal modeling. In particular, the WTD model in MAS has allowed us to better capture the complex interplay between magnetic fields, plasma dynamics, and energy transport in the solar corona, leading to more reliable output of solar wind conditions and coronal structures \citep{2016ApJ...832..180D,2018NatAs...2..913M,2018ApJ...856...75T,2019SoPh..294...13L,2021A&A...650A..19R,2023ApJ...959...77L,2023ApJ...959L...4M}. 

Second, the model is driven by a single Carrington rotation (CR) map. Thus, although the model is advanced in time, the output represents a steady-state equilibrium of the structure of the corona during the interval defined by the CR. This is a reasonable approximation when not investigating transient features specifically, such as coronal mass ejections (CMEs), or smaller-scale phenomena like jets or magnetic switchbacks \citep{Bale2019} that are a key driver of solar wind evolution \citep{Halekas2023, Rivera2024}. Recently, we presented a new paradigm for modeling coronal structure using a time-evolving synchronic photospheric magnetic field map \citep{2023ApJ...959...77L,2023ApJ...959L...4M,downs24a}. However, while promising, it remains a ``proof-of-concept'' approach that requires substantial computational resources to undertake and is beyond the scope of the present study. 

Third, while the corona and heliosphere are separated as distinct regions, the output of the coronal solution is used as direct input into the heliospheric model, producing a seamlessly coupled description of the solar wind plasma from 1 $R_S$ to 1 AU \citep{2013ApJ...777...76L}. As described in more detail in Section~\ref{cs-desc}, the key information from the mid-corona outward is the plasma velocity field, which varies smoothly across the coronal-heliospheric boundary. 

For the coronal simulation, we prescribe  the surface magnetic flux at $r=R_\sun$ from  smoothed MDI
magnetic field synoptic charts for Carrington rotations 2002 and 2063.
The  maps are given
 as input to the MHD WTD model \citep{2018ApJ...856...75T,2018NatAs...2..913M}. 
For the  latter, we use a nonuniform 
grid in $r\times\theta\times \phi$ of $269\times181\times361$ points
extending from $R_\odot$ to $30 R_\odot$.
The smallest radial grid spacing
at $r=R_\sun$   was   $\sim400$ km; the    angular resolution in $\theta$
ranged between $0.8^\circ$ at the equator and $1.7^\circ$ at the poles;
the $\phi$ mesh was uniform.
To dissipate structures  that are  smaller than the cell size 
and cannot be resolved,
we prescribe a uniform resistivity $\eta$ corresponding to a
resistive diffusion time $\tau_R \sim 4\times 10^4$ hours, which
is much lower than the value in the solar corona. 
 The Alfv\'en
travel time at the base of the corona ($\tau_A=R_\sun/V_A$) for
$|\mathbf{B}|=2.205~\mathrm{G}$  and $n_0=10^8~\mathrm{cm}^{-3}$, which
are typical reference values,
is 24 minutes (Alfv\'en speed $V_A= 480~\mathrm{km/s}$),
so the Lundquist number  $\tau_R/\tau_A$ was $1\times 10^5$.
We introduce 
a uniform
viscosity $\nu$, corresponding to a
viscous diffusion time $\tau_\nu$ such that $\tau_A/\tau_\nu=0.015$, in order to dissipate unresolved scales
without substantially affecting the global solution.
We prescribed fixed chromospheric values
 of density and temperature at the base of the domain of $n_0=4\times
10^{12}~\mathrm{cm^{-3}}$ and $T_0=17{,}500~\mathrm{K}$, respectively.
These values were set to form a chromospheric   ``temperature plateau''
that remains sufficiently large \citep{2009ApJ...690..902L} during
the calculation no matter how large the heating is.

For the coronal heating term, we use the same WTD model parameters as the simulation described in \citet{2023ApJ...959...77L}, which is close to the numbers used in  \citet{2018NatAs...2..913M}. The Poynting flux of wave energy is prescribed at the base of the corona
through an amplitude of the Els\"asser variable 
$z_0=9.63~\mathrm{km/s}$, and we set the transverse correlation scale $\lambda_0
= 0.02 R_\odot$ along with a scaling factor $B_0=8.53~\mathrm{G}$ such that
$\lambda_\perp= \lambda_0 \sqrt{B/B_0}$ in the corona.
Similar to \citet{2018NatAs...2..913M} in adding two small
exponential heating terms to heat the low corona: 
$H_0=2.7\times 10^{-5}~\mathrm{erg/cm^3/s} $, $\lambda_0 = 0.03 R_\odot$; 
$H_0=1.6\times 10^{-8}~\mathrm{erg/cm^3/s} $, $\lambda_0 =  R_\odot$.
Likewise,  the 
wave pressure was specified from the WKB model \citep{2009ApJ...690..902L}.

We start the coronal calculations using the magnetic flux maps
to calculate potential field extrapolations. The plasma
temperature, density, and velocity are imposed from a 1D solar wind solution that has been calculated previously.
Then, we advance the MHD equations for about 80 hours to relax the
system to a steady state.

The heliospheric model extends through 
 $25 R_\sun \leq r \leq 230 R_\sun$ and is discretized on a non-uniform
 $401\times181\times361$ mesh in $(r,\theta,\phi)$. The radial and
longitudinal resolution is
constant,  the  minimum and maximum sizes of the latitudinal mesh
are the same as in the coronal model.
 We have prescribed a uniform resistivity profile, such
that the ratio of the resistive dissipation time with the Alf\'en
wave propagation time is $\tau_R/\tau_A=10^6$. 
We have also introduced
a uniform
viscosity $\nu$, corresponding to a
viscous diffusion time $\tau_\nu$ such that $\tau_A/\tau_\nu=0.001$.

Using a set of fields extracted at $r=25 R_\odot$ from the relaxed
coronal calculation, we have  specified
a potential extrapolation
as initial condition for the magnetic field of the heliospheric
model. We have radially extrapolated 
the values of $v_r$, $\rho $, and $T$
to initialize the plasma properties as shown in \citet{2013ApJ...777...76L},
as well as the charge states.
Then we advanced the MHD equations with fixed boundary values
 for 161 hours to ensure that a steady-state solution is formed in the heliosphere.


\subsubsection{Time-Dependent Charge State Model}
\label{cs-desc}

The process by which the state of ionization of a particle in the solar corona gets ``frozen-in'' is a complex interplay between ionization rates, recombination rates, and the density of the expanding solar corona. Ionization Rates are the rates at which neutral atoms or ions collide with free electrons to become more highly charged ions (i.e., lose additional electrons). They depend on the electron temperature and the density of the solar corona, with higher temperatures and densities generally increasing rates.
Recombination Rates, on the other hand, are the rates at which ions capture free electrons to become less charged (i.e., gain electrons). They also depend on the electron temperature and density, with higher temperatures typically decreasing recombination rates, while higher densities increase them.
With density decreasing substantially with distance from the solar surface within a relatively isothermal corona, it affects both ionization and recombination rates significantly and effectively defines the ``freeze-in'' point where the density is too low for ionization and recombination to keep up with the changing conditions - the density becomes so low that the timescales for ionization and recombination processes become much longer than the expansion timescale of the solar wind. 

\begin{figure}[ht!]
    \centering
    \includegraphics[width=0.975\textwidth]{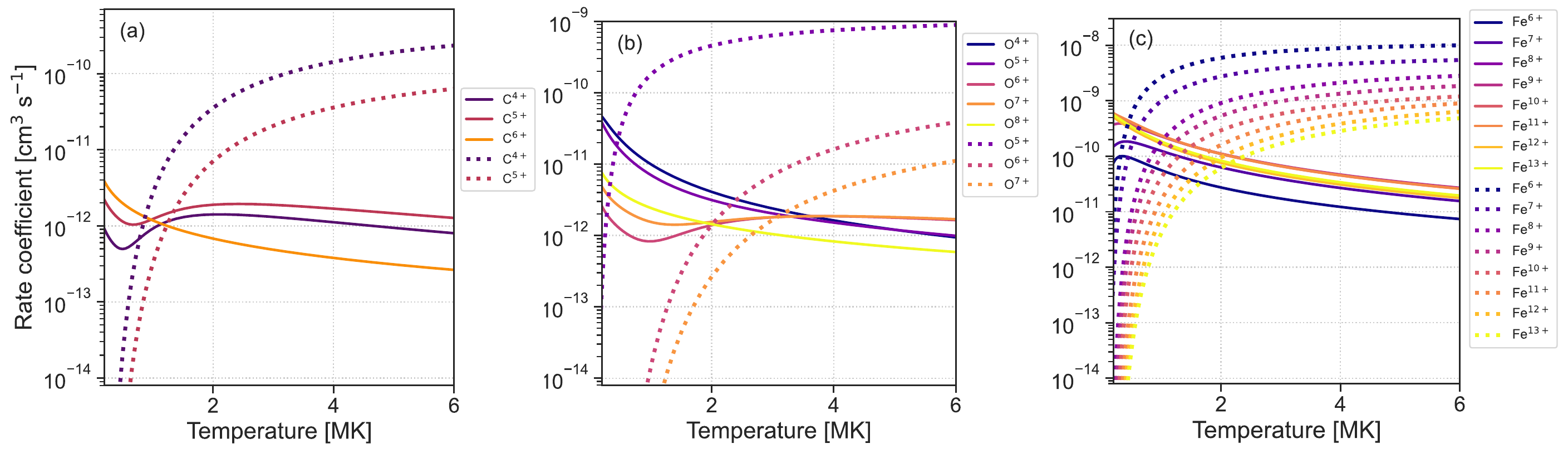}
    \caption{Ionization and recombination rates for (a) carbon, (b) oxygen, and (c) Iron where the dotted lines represent the ionization and solid lines represent the recombination rate coefficients for the individual ions.   
}
    \label{ion-recomb}
\end{figure}

In Figure~\ref{ion-recomb}, we show the rate coefficients for (a) carbon, (b) oxygen, and (c) iron as a function of temperature computed using atomic parameters from CHIANTI v10 \citep{delzanna2021}. The ionization rate coefficients are shown by the dotted lines, while the recombination rate coefficients are shown by the solid lines. Where the two lines cross (that is, lines of the same color), this represents the temperature at which the two processes of ionization and recombination match. Moving away from the Sun along field lines, the temperature decreases, and beyond these crossovers, recombination rates dominate. At the same time, however, the density decreases substantially such that, at some point ($n < 600$ cm$^{-3}$), the timescale for either process becomes longer than the transit time of the particle through the corona. 

Under equilibrium conditions, we can represent this process with the following simple ionization balance equation:
\begin{equation}
\frac{dN_{i+1}}{dt} = n_e C_i(T) N_i - n_e \alpha_{i+1}(T) N_{i+1} \label{eq_cs_simple}
\end{equation}
where \( N_i \) is the number density of ions in charge state \( i \), \( N_{i+1} \) is the number density of ions in charge state \( i+1 \), \( n_e \) is the electron density, \( C_i(T) \) is the ionization rate coefficient for ions in charge state \( i \), \( \alpha_{i+1}(T) \) is the recombination rate coefficient for ions in charge state \( i+1 \), and \( T \) is the electron temperature. 

At the freeze-in point, the rate of change of the ionization state becomes negligible, so the left-hand side of the equation can be approximated to zero, and we get the equilibrium condition:

\[
C_i(T) N_i \approx \alpha_{i+1}(T) N_{i+1},
\]

or: 

\[
  N_{i+1} / N_i \approx C_i(T) / \alpha_{i+1}(T) 
\]

Thus, in principle, knowledge of $N_{i+1} / N_i$, together with the information from Figure~\ref{ion-recomb}, should allow us to infer the temperature of the plasma that produced the observed charge state ratios, effectively providing yet another crucial constraint for global scientific models. 

The present version of MAS  incorporates a
 non-equilibrium ionization module to advance the
 fractional charge states of minor ions according to the model
of \citet{2015A&C....12....1S}:
\begin{equation}
\frac{dN_{i+1}}{dt}=
\frac{\partial N_{i+1}}{\partial t} + \mathbf{v} \cdot \nabla N_{i+1} =
n_e \left [ C_{i} N_{i} - \left ( C_{i+1} + \alpha_{i+1} \right ) N_{i+1} + \alpha_{i+1}  N_{i+2}  \right ]. \label{e:cs}
\end{equation}
Equation~(\ref{e:cs}) also accounts for ionization and recombination for the $i+2$ charge state, which are safely neglected in Eq.~(\ref{eq_cs_simple}) near equilibrium conditions \citep{lecointre13a}.
The rate coefficients are
derived
from the CHIANTI atomic database
\citep{1997A&AS..125..149D,2013ApJ...763...86L}. 
$\mathbf{v}$ indicates the velocity of the plasma.
This module has
been employed in 
 3D calculations of the corona \citep{2023ApJ...955...65R,2023ApJ...959...77L}.

\section{Results}

\subsection{Model Validation}

Before comparing modeled and observed charge states, it is worth validating the model results using emission images from a variety of wavelengths. These measurements are particularly sensitive to the temperature and density profiles in the low corona and serve as an independent ``ground truth'' for the modeled plasma and magnetic field parameters. Coronal holes (dark), active regions (structured and bright), and quiet sun regions (dim) can be compared by eye to assess -- at least qualitatively -- to what extent the model is consistent with the observations. 

\begin{figure}[ht!]
    \centering
    \includegraphics[width=0.95\textwidth]{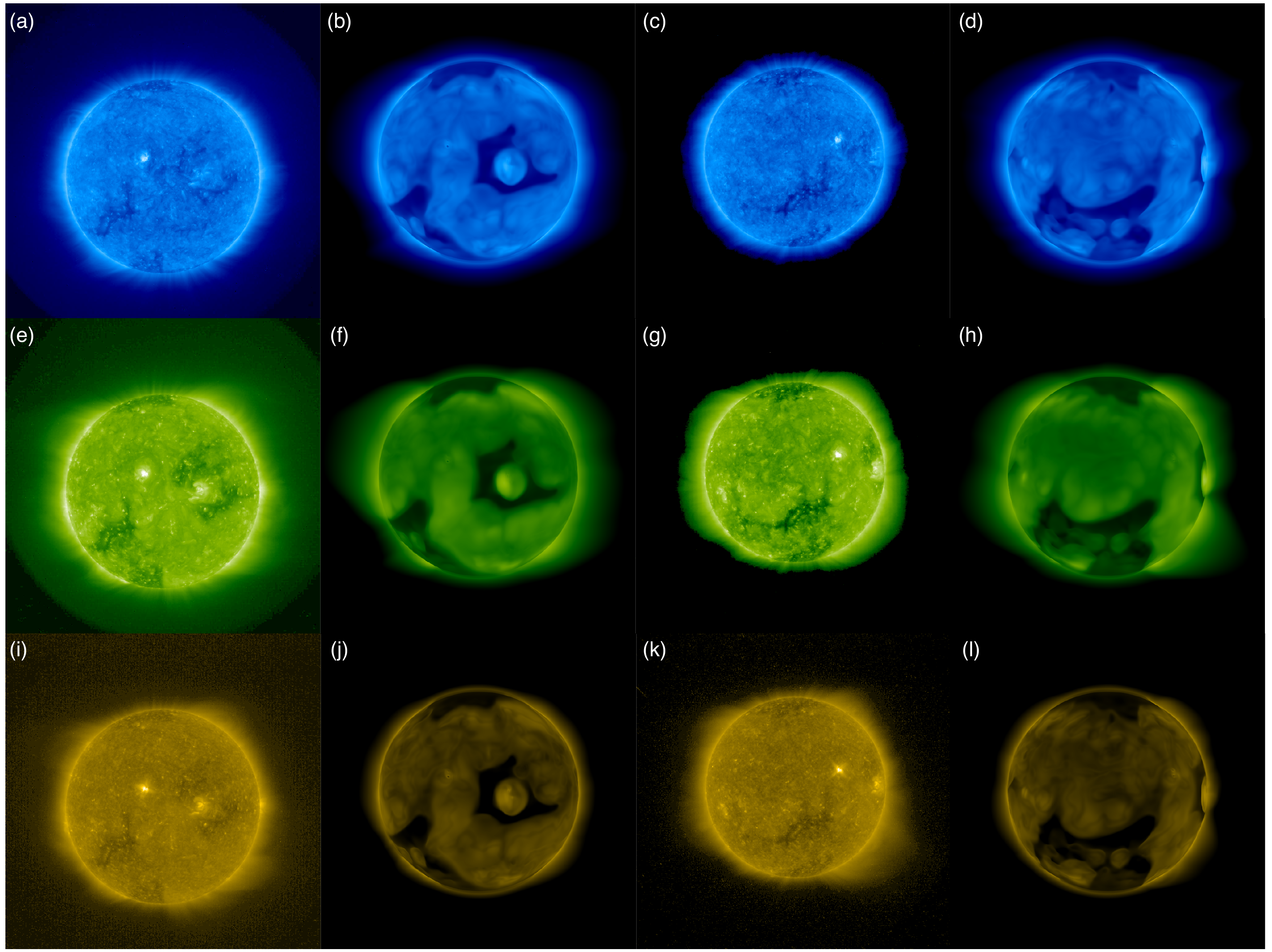}
    \caption{Comparison between observed (a, c, e, g, i, and k) and modeled (b, d, f, h, j, and l) logarithmic emission for STEREO EUVI 171 \AA, 195 \AA, and 284 \AA\ for CR2063. STEREO-A comparisons are in the first two columns and STEREO B are in the last two columns.}
    \label{emission-comp-2063}
\end{figure}

Figure~\ref{emission-comp-2063} compares observed (a, c, e, g, i, and k) and modeled (b, d, f, h, j, and l) logarithmic emission for STEREO EUVI 171 \AA, 195 \AA, and 284 \AA\ for CR2063. STEREO-A comparisons are in the first two columns and STEREO B are in the last two columns. These narrowband images provide strong constraints for the models since the amount of intensity (or lack thereof) depends sensitively on the temperature profiles in the corona, which, in turn, are intimately tied to the coronal heating profiles. The overall intensities are reasonably well-matched between observations and simulation, suggesting that the basic heating parameters are reasonable; however, it should be noted that in general, the model intensities are slightly less than the observed values, particularly at 284 \AA. The location, size, and relative orientation of the coronal holes are well-matched by the model at both viewpoints. The circular coronal hole at STEREO A near the equator is more prominent in the model than in the observations, as is the broad southern coronal hole at STEREO B. 

\begin{figure}[ht!]
    \centering
    \includegraphics[width=0.6\textwidth]{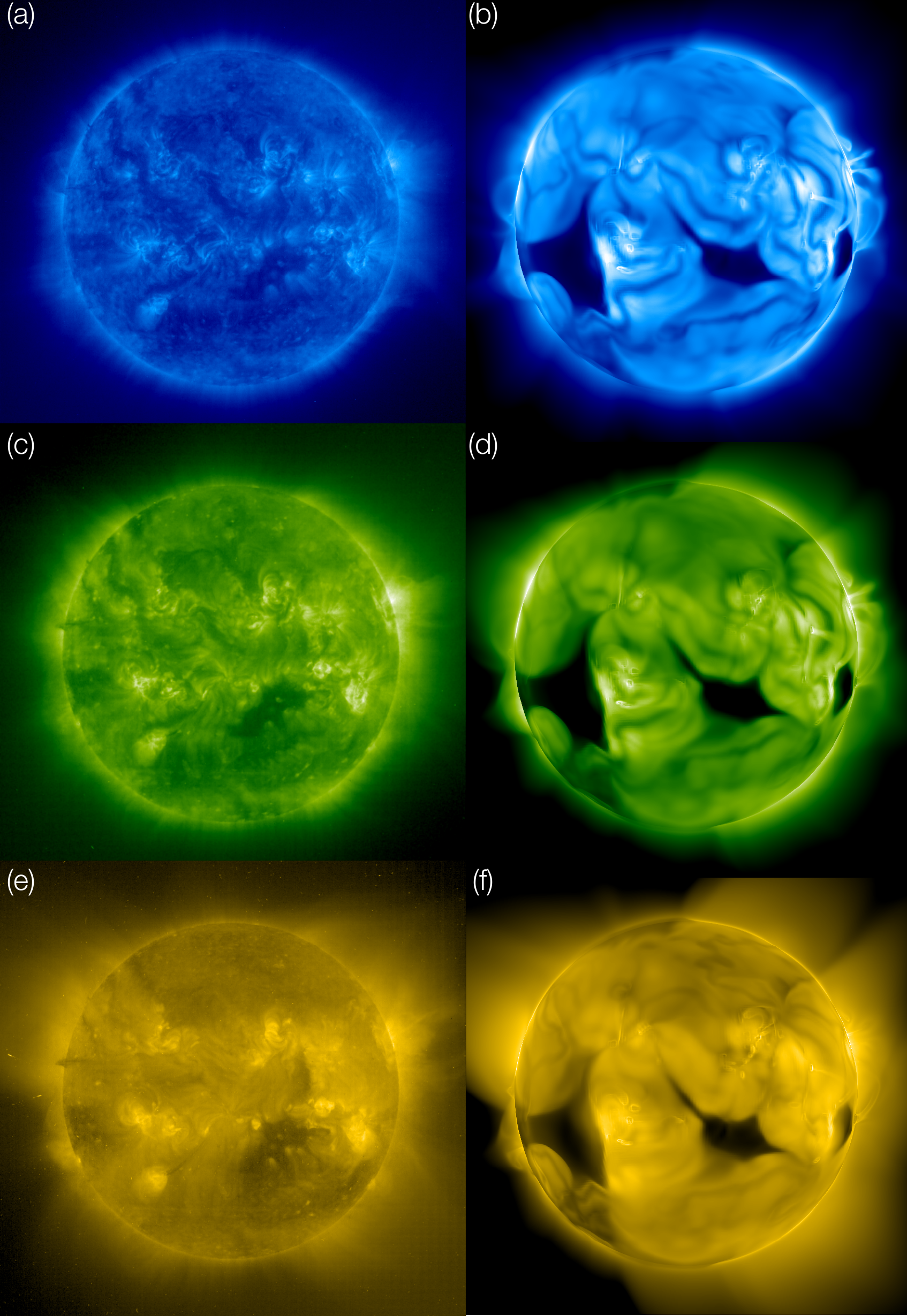}
    \caption{Comparison between observed (a, c, and e) and modeled (b, d, and f) logarithmic emission for SOHO's (a) EIT 171 \AA, (b) EIT 195 \AA, and (c) EIT 284 \AA\ for CR2002. The Carrington longitude of the observer was set to $180^\circ$, corresponding to 08:36 UT on 2003/04/29.}
    \label{emission-comp-2002}
\end{figure}

Figure~\ref{emission-comp-2002} compares observed (a, c, and e) and modeled (b, d, and f) logarithmic emission for the Solar and Heliospheric Observatory (SOHO)  (a) EIT 171 \AA, (b) EIT 195 \AA, and (c) EIT 284 \AA\  for CR2002. The Carrington longitude of the observer was set to $180^\circ$, corresponding to 08:36 UT on 2003/04/29. We note here that for CR2002 the overall intensities are well-matched between observations and simulation, suggesting that the basic heating parameters are reasonable. The location, size, and relative orientation of the two major equatorial coronal holes are not particularly well reproduced by the model. The one to the east was not as well developed, while the shape of the one to the west differs from the observations. Given the more rapid evolution of the underlying field during this phase of the solar cycle, such differences are not surprising. It is interesting to note that the bright loops observed above an AR on the west limb do appear to be particularly well matched in the model, as the approximate locations of the ARs on the disk. Finally, several of the filament channels are captured well in the model, such as the one running from the NE limb to disk center, as seen in 171 and 195 \AA. 

\subsection{Model Output}

\begin{figure}[ht!]
    \centering
    \includegraphics[width=0.666\textwidth]{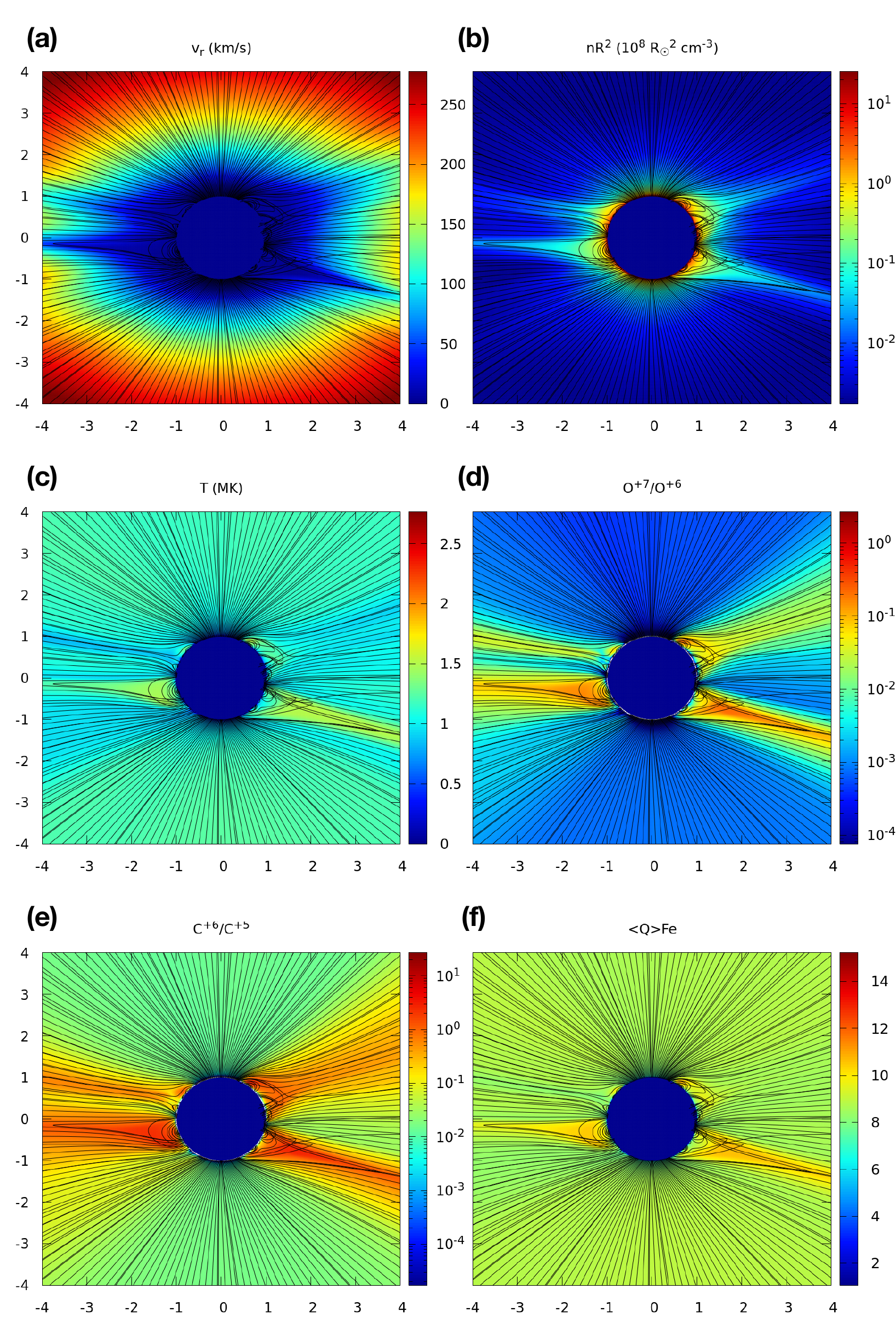}
    \caption{Meridional slices of (a) radial velocity, (b) plasma density, (c) temperature, (d) $O^{7+}/O^{6+}$, (e) $C^{6+}/C^{5+}$, and (f) the average charge state of Fe, at Carrington longitude 180$^{\circ}$, for CR 2063. A selection of magnetic field lines have been traced both from the inner and outer boundaries, which were, at least initially, in the plane of the image. 
}
    \label{vnT_cs_mer_fl-2063}
\end{figure}

In addition to creating model output products that mimic what remote solar observatories measure, such as white-light or emission images, the models allow us to construct and combine a range of parameters that are not directly observable. In Figure~\ref{vnT_cs_mer_fl-2063}, for example, we compare meridional slices of radial velocity, plasma density, temperature, average charge state of Fe, $O^{7+}/O^{6+}$, and $C^{6+}/C^{5+}$ at Carrington longitude 180$^{\circ}$, for CR 2063. On top of each, a selection of magnetic field lines has been drawn (both outwardly and inwardly), which were, at least initially, in the plane of the image.

Overall, the panels convey the late declining phase structure of the corona that was prevalent during the approach to the 2009 minimum \citep{riley11a}. This minimum, unlike the previous one, had what at the time was considered to be a rather unique streamer structure, but given the similarity of the most recent minimum in 2019, it might be thought of as a new `normal' state. Instead of the (apparently)  single equatorial streamer belt observed in earlier and stronger cycles (that is, in coronagraph images, two streamers – one emanating from the eastern equatorial limb of the Sun and the other from the West), the streamer structure was more complex. In the images shown here, while the polar regions are still filled with fast, hot, and tenuous plasma, the equatorial regions are composed of a pair of streamers emanating from each limb. As noted by \citet{riley11a}, this was likely due to the reduced amplitude of the polar fields as compared to the minimum in 1996, say. The distribution of plasma is consistent with our understanding of streamers and the slow-speed wind that is associated with them. Within the closed field regions, the plasma is dense, stationary, and hot. At the edges, it is slow-moving and dense. The composition data is consistent with this. The average charge state of iron shows moderately elevated values of 10 within the closed streamer loops, dropping to 8 along open field lines. It is slightly elevated along the open field lines close to the closed loops, where the wind is slower and denser. The $O^{7+}/O^{6+}$ and $C^{6+}/C^{5+}$ ratios, in contrast, maintain a ``beam'' of elevated values stretching out from the base of the streamer. Closest to the streamers, these values are higher, but the key point is that, unlike the density, speed, temperature, and $<Q_{Fe}>$ parameters, which pinch down as the streamers do, the $O^{7+}/O^{6+}$ and $C^{6+}/C^{5+}$ maintain a constant width. The centroid of this swath, however, does not follow a radial trajectory but tracks along the direction of the streamer, cusp, and presumably current sheet location, if present. 

\begin{figure}[ht!]
    \centering
    \includegraphics[width=0.666\textwidth]{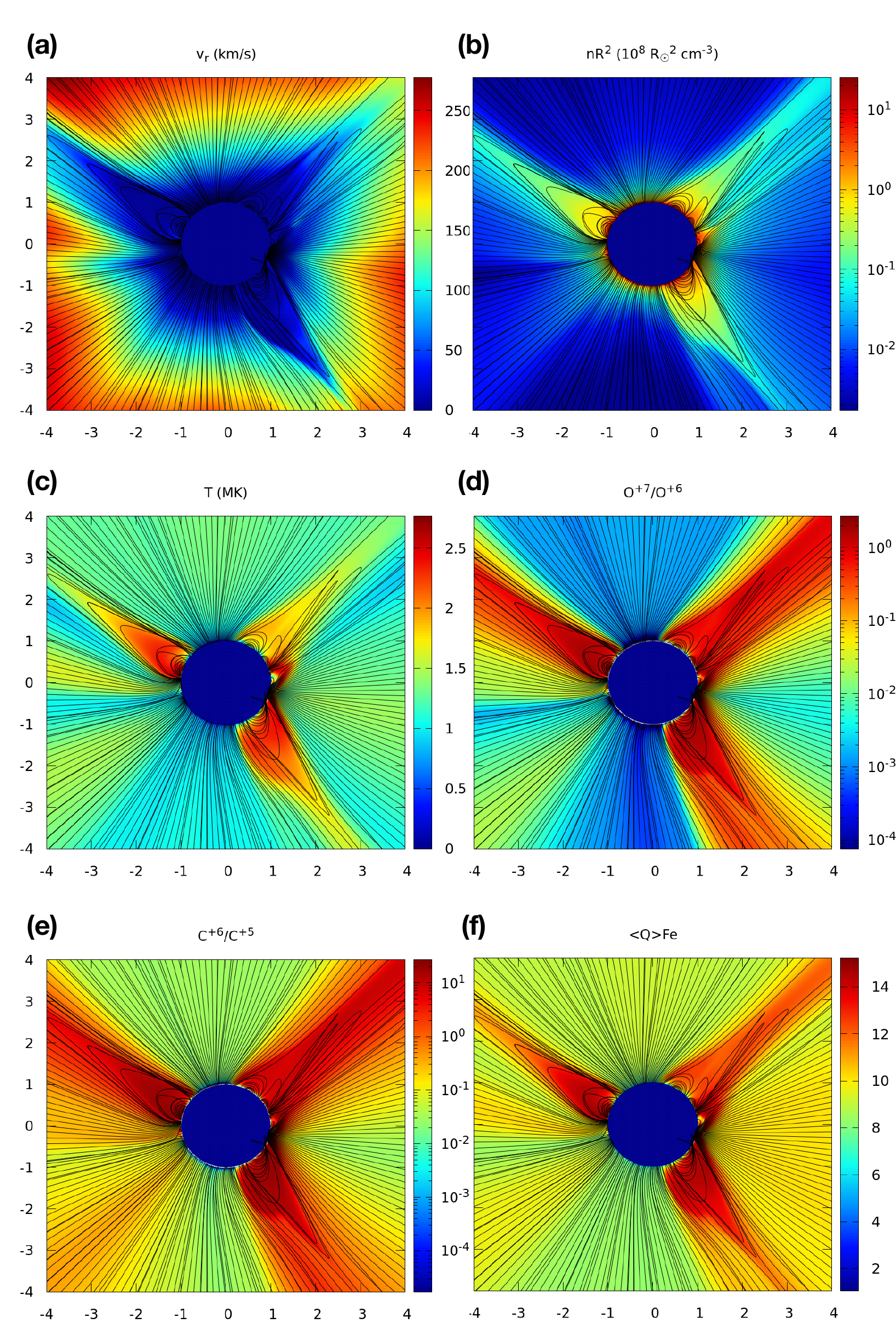}
    \caption{As Figure~\ref{vnT_cs_mer_fl-2063}, but for CR2002.  
}
    \label{vnT_cs_mer_fl-2002}
\end{figure}

In Figure~\ref{vnT_cs_mer_fl-2002}, we show the same parameters for CR2002. This slice was chosen since it captures a solar maximum-like configuration. The N-E and S-W quadrants contain standard dipolar streamers, where a current sheet continues off their tips, separating the oppositely directed fields. In the N-W quadrant, however, a quadrupolar geometry sets up a pseudo-streamer (which can be seen by the double loops within the streamer). The field lines bounding this structure are both of the same polarity. 

We note the following points. First, within closed field line regions, the plasma is dense, static, and hot. Additionally, the values of $O^{7+}/O^{6+}$ and $C^{6+}/C^{5+}$, as well as the average charge state of Fe, are elevated. Interestingly, along the boundaries of the closed field lines, while the plasma parameters change abruptly, $O^{7+}/O^{6+}$ and $C^{6+}/C^{5+}$ values appear to remain high, that is, there is a swath of high charge state ratios mapping out from the full meridional extent of the base of the closed field lines. While this may be, at least in part, an artifact of these parameters being presented logarithmically, the quality of the profiles is different between the charge state ratios and the plasma parameters. Take, for example, the yellow-speed contour in the southeast quadrant of panel (a). It is effectively vertical, suggesting that the speed decrease from the equator to the pole is sensitive to height. This makes intuitive sense as we are traversing from the center of a coronal hole to the tip of a streamer, and fast wind accelerates to its final, asymptotic speed more quickly than slow wind. In contrast, the analogous yellow contour for $O^{7+}/O^{6+}$ is orientated closer to 45$^{\circ}$, such that the variation from the equator to the pole is independent of height. Or, that the charge-state ratio is frozen in the low corona and mapped out into the upper corona by the radial plasma flow. Interestingly, the average charge state of Fe maps outward more like the plasma profiles than the charge-state ratios, suggesting that it is not frozen as low in the corona. 

Comparison of the profiles in Figure~\ref{vnT_cs_mer_fl-2002} with those in Figure~\ref{vnT_cs_mer_fl-2063} highlights some significant differences between the solar minimum and maximum properties of modeled charge-state profiles. Although the speed and density variations are not that dissimilar, the temperatures within the closed loops at solar maximum (CR2002) are substantially higher. This alone provides a natural explanation for the substantially larger charge state ratios and elevated Fe ionization seen in panels (d)-(f) of Figure~\ref{vnT_cs_mer_fl-2002}. The model then predicts that, at least, surrounding the heliospheric plasma sheet, the increase in these parameters should be substantially greater during more active times than near the minimum of the solar cycle.  

A final point worth remarking upon concerns the asymmetries in moving across the streamers and pseudo-streamer. Consider, for example, the streamer in the S-W quadrant again. Note how the latitudinal gradient at a particular height is much stronger moving towards the south pole than it is moving towards the equator. This can be seen in $O^{7+}/O^{6+}$ and $C^{6+}/C^{5+}$ (panels (d) and (e)) but also in the scaled density (panel (b)). To lesser degrees, the other stream and pseudo-streamer also show asymmetries.

\begin{figure}[ht!]
    \centering
    \includegraphics[width=0.666\textwidth]{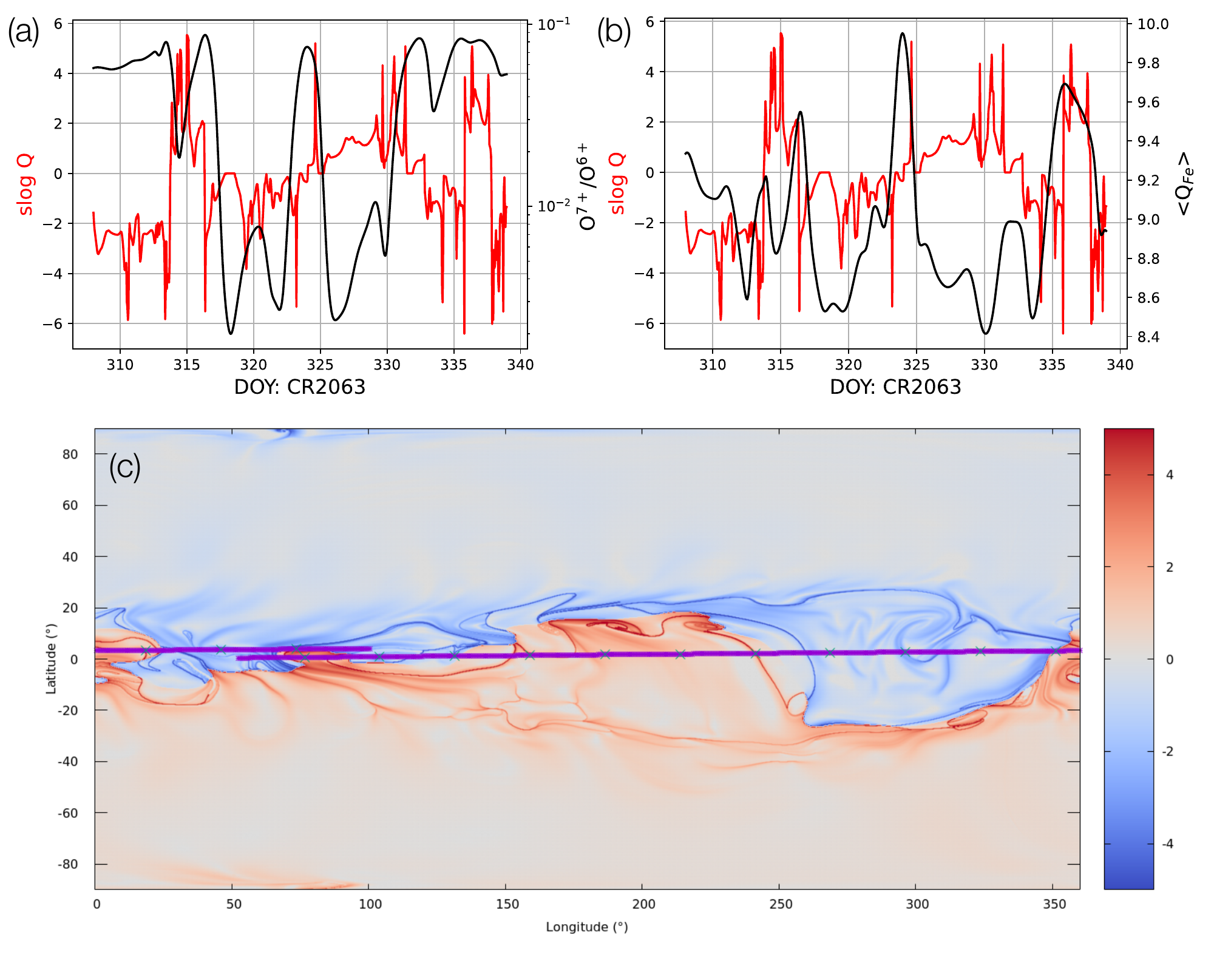}
    \caption{Comparison of slogQ with (a) $O^{7+}/O^{6+}$ and (b) $<Q_{Fe}>$ for CR2063. slogQ is shown in red, and the charge information is shown in black. In (c), the full map of slogQ is shown, together with the trajectory of the ACE spacecraft (purple). Note that the trajectory starts at $\sim 100^{\circ}$ longitude, moves to earlier Carrington longitudes (left), crosses the 0/360 boundary, and continues from the right of the panel moving left.   
}
    \label{slog-cr2063}
\end{figure}

\begin{figure}[ht!]
    \centering
    \includegraphics[width=0.666\textwidth]{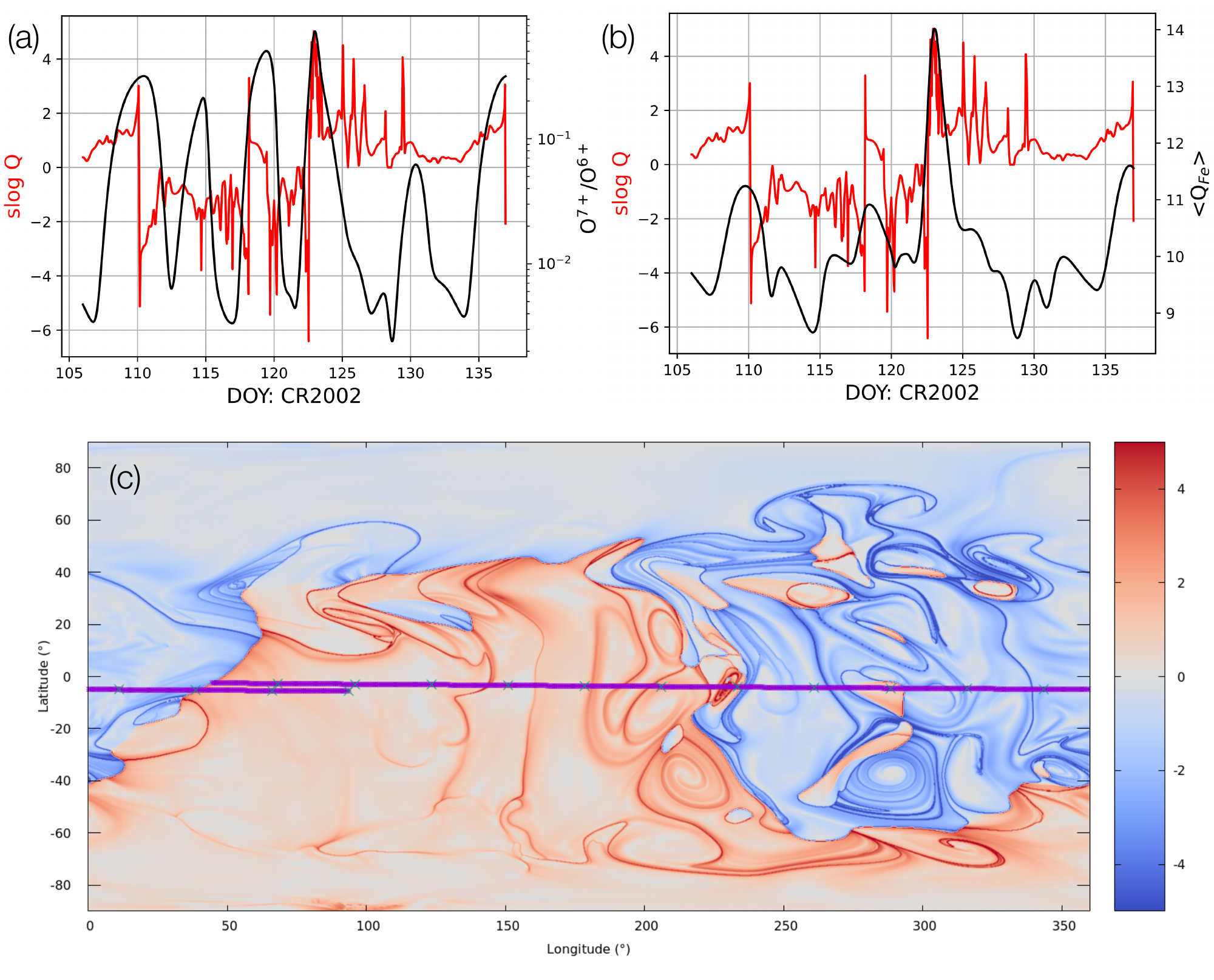}
    \caption{As Figure~\ref{slog-cr2063}, but for CR2002.   
}
    \label{slog-cr2002}
\end{figure}

The magnetic field structure inferred from the MHD model can help us understand the underlying physical processes at work. The primary tool for such analysis is the 
`squashing factor' $Q$ \citep{titov02a,titov07a}. 
High-$Q$ regions reveal separatrix surfaces and quasi-separatrix layers (QSLs), which fully or partly partition magnetic fields into different flux systems. Surprisingly, even open magnetic fields in realistic models of the corona have very complex 
structure known as the S-web \citep{antiochos11a,linker11a,titov11a}. We can connect the patterns observed in Figure~\ref{vnT_cs_mer_fl-2063} with the underlying magnetic structure by exploring the signed variation of Q. Figure~\ref{slog-cr2063} compares slogQ with (a) $O^{7+}/O^{6+}$ and (b) $<Q_{Fe}>$ for CR2063. The trajectory of the ACE spacecraft, which is shown in purple) starts at $\sim 100^{\circ}$ longitude, moves to earlier Carrington longitudes (left), crosses the 0/360 boundary, and continues from the right of the panel moving left. Focusing first on slogQ, we note that the HCS is confined to a band of within 20/30$^{\circ}$ about the equator in the northern/southern hemisphere, respectively. Additionally, the blue and red arcs that branch off from the slogQ=0 contour suggest the presence of pseudo-streamers. In terms of what is sampled by ACE, the four sector crossings are suggestive of a complex magnetic topology despite the relative simplicity of the structure beyond the equatorial band. This can be contrasted with the declining phase of cycle 22, during which time a simpler tilted dipole geometry with a simple two-sector pattern was observed by both Ulysses and ACE \citep{gosling95f,1996JGR...10124349R}. Comparisons between $O^{7+}/O^{6+}$ and slogQ show that when the latter changes sign (at the HCS, and, by inference, the heliospheric plasma sheet) $O^{7+}/O^{6+}$ peaks (e.g., DOY 316, 324, 332, and 336). This is also true for $<Q_{Fe}>$, although the relative peaks change significantly from one to another. Additionally, we note that the peaks in both $O^{7+}/O^{6+}$ and (b) $<Q_{Fe}>$ are relatively symmetric; that is, the rise time is approximately the same as the decay time. 

Charge state and squashing factor are also compared for CR2002 in Figure~\ref{slog-cr2002}. 
Focusing on the $O^{7+}/O^{6+}$ and $<Q_{Fe}>$ variations first, we note that the peak profiles are this time asymmetric with sharp rises in the charge state ratios coinciding with sector boundary crossings (e.g., DOY 122.5) and more shallow decreases associated with drops in, but no change in sign of slogQ. We note further that the asymmetric peaks can be shallow/steep (e.g., DOY 119) or steep/shallow (DOY 123). Panel (c) allows us to interpret the streamer structure seen in Figure~\ref{vnT_cs_mer_fl-2002}. Since that view was from 180$^{\circ}$ longitude, the east and west limbs correspond to longitudes of 90$^{\circ}$ and 270$^{\circ}$ in panel (c) of Figure~\ref{slog-cr2002}, respectively. For the former, we see that moving from -90$^{\circ}$ to +90$^{\circ}$ we cross from positive to negative polarity at $\approx 40^{\circ}$ latitude, corresponding to the latitude of the streamer. For the latter, we cross from positive to negative polarity at $\approx -65^{\circ}$ latitude, remaining in negative polarity essentially all the way to the north pole, except where we clip a positive polarity island at $\approx 45^{\circ}$ latitude, corresponding to the location of the pseudo-streamer.

\subsection{Comparative Analysis}

We now turn our attention to a more detailed comparison of the observed and modeled charge states of key ions (e.g., $O^{7+}/O^{6+}$, $C^{6+}/C^{5+}$, and the average charge state of iron $<Q_{Fe}>$). We start by comparing the full charge states of C, O, and Fe as a function of time through each Carrington rotation, as well as the plasma and magnetic field variations. We then consider the statistical properties of the parameters. 

\begin{figure}[ht!]
    \centering
    \includegraphics[width=0.95\textwidth]{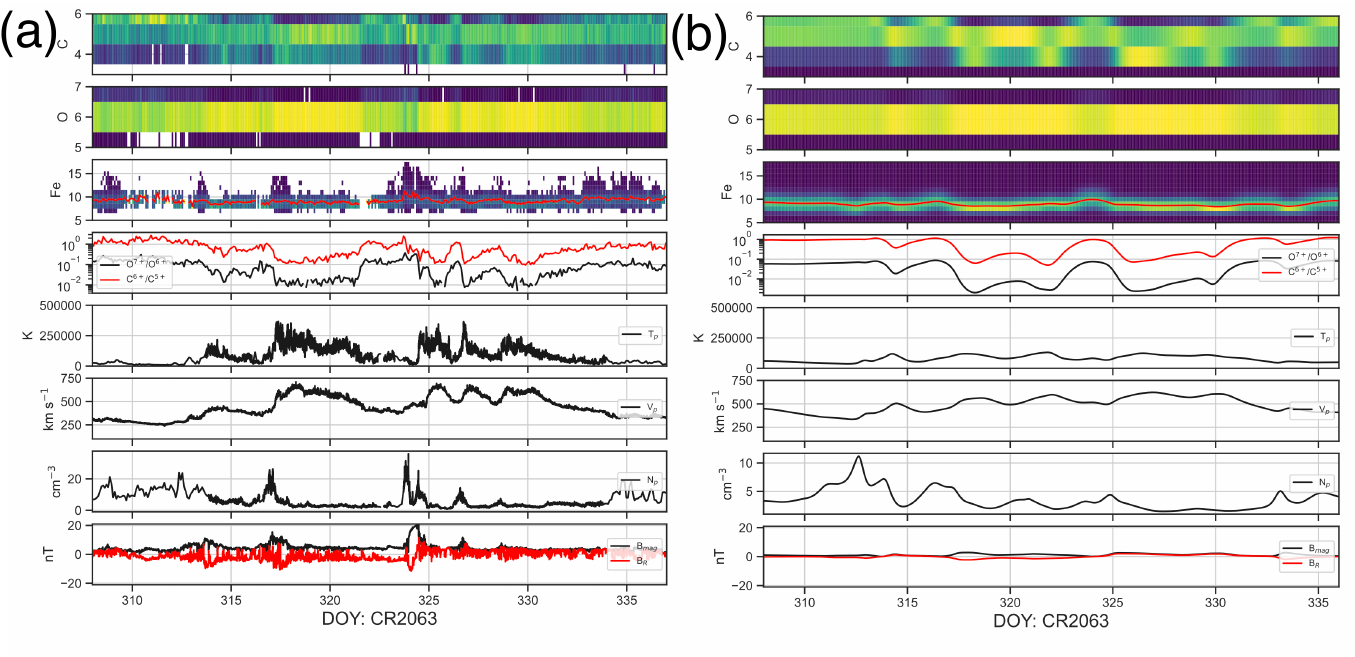}
    \caption{Comparison between observed (a) and modeled (b) in-situ measurements at ACE for CR2063, corresponding to November 04 to December 01, 2007 (doy: 308 to 335). The top three panels show heat maps of the charge state distributions of C, O, and Fe, respectively, with the average charge state of Fe marked with the red curve. The next panel shows the ratio of $O^{7+}/O^{6+}$ (black) and $C^{6+}/C^{5+}$ (red). The remaining panels show proton temperature, proton speed, proton density, and magnetic field (radial component (red) and magnitude (black)). 
}
    \label{ace-cr2063-obs-sim}
\end{figure}

In Figure~\ref{ace-cr2063-obs-sim} we compare the observed (a) and modeled (b) composition and plasma values during CR2063. An ICME was observed during days 323-324 (2007-11-19 and 2007-11-20), which was not captured in the model. We note that the y-axis scales in each panel (with the exception of number density) have been scaled identically allowing us to make -- at least qualitatively -- direct comparisons with the values. Using this comparison, we focus more on a statistical comparison between the two. We note several points. 
First, for C, O, and Fe, we see a generally positive match between observations and model results. The dominant charge states, their spread amongst different states, and even the temporal variability appear to match well. For example, the Fe charge states are narrowly constrained about nine, O is almost entirely at six, and C is predominantly set at five, but has intervals where the ratio is reversed, with six dominating. These ratios are captured more explicitly in the fourth panel, which shows traces of $O^{7+}/O^{6+}$ (black) and $C^{6+}/C^{5+}$ (red). The relative position of the ratios is matched between model and observation, as is the overall evolution of the profiles during the interval. One notable disagreement, perhaps, is that the relative offset of the two traces is larger in the model than in the observations. However, at least to some extent, this may be an artifact of high-frequency oscillations in the data, which may be noise or a real physical process that is not captured by the model, which gives the impression that the two profiles are closer together. Comparisons of the plasma (temperature, speed, and density), as well as magnetic field (radial component and amplitude), show qualitative matches, but also some obvious disagreements. Of particular note is that the variations in the parameters do not appear to be as large in the model results. We will return to this issue later. 

\begin{figure}[ht!]
    \centering
    \includegraphics[width=0.95\textwidth]{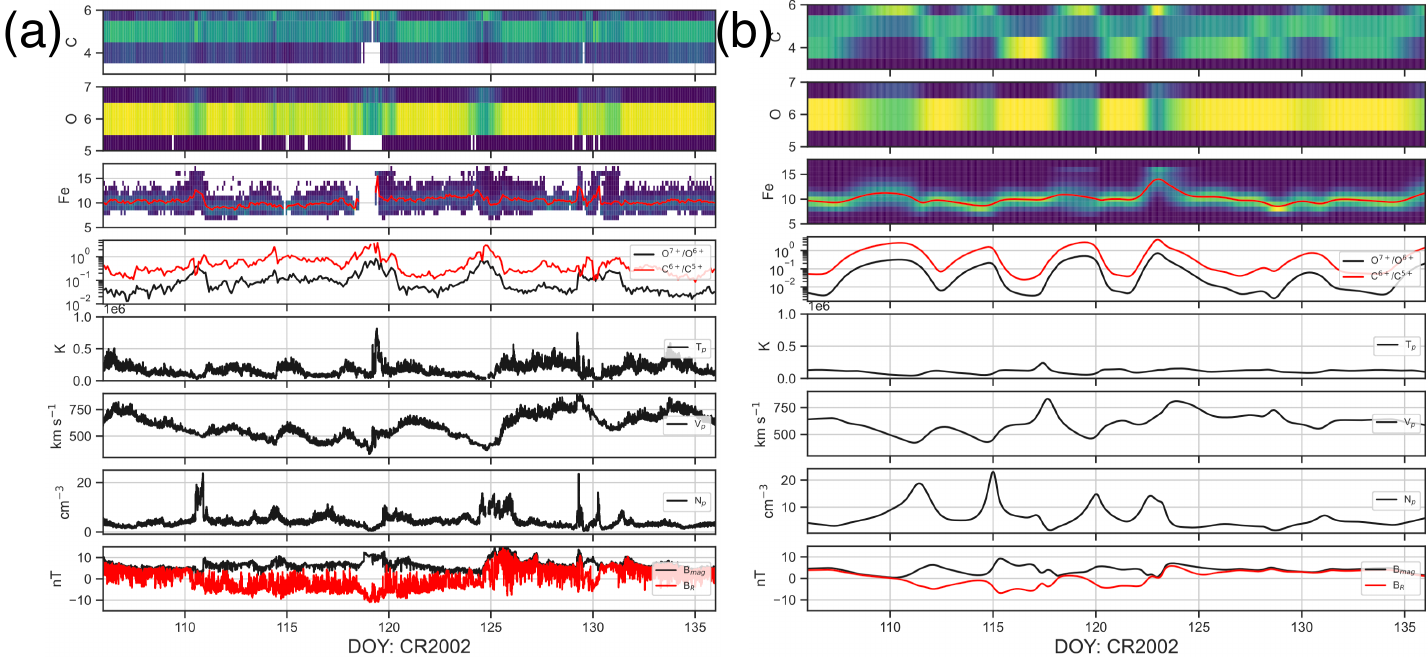}
    \caption{As Figure~\ref{ace-cr2063-obs-sim}, but for CR2002, corresponding to April 16 to May 16, 2002 (doy: 106 to 136). 
}
    \label{ace-cr2002-obs-sim}
\end{figure}

In Figure~\ref{ace-cr2002-obs-sim}, we make an analogous comparison for CR2002. 
The first and most obvious point to note is that the absolute values for C, O, and Fe in the heatmaps (top-three panels) are considerably larger for CR2002 than for CR2063. 
Again, the overall values and distributions captured in the heatmaps are well-matched by the model. The predominant charge states of 5+, 6+, and 10+ for C, O, and Fe respectively agree well, as does the distribution of states (vertical spread) for each atomic species. Second, as ratios, the $O^{7+}/O^{6+}$ (black) and $C^{6+}/C^{5+}$ (red) averages and variability also track well. Third, while the average proton temperatures are not dissimilar between the observations and model, the model does not reproduce the occasional peaks, due to the compression and heating of plasma as faster wind attempts to overtake slower wind. Fourth, speed variations, which range from 400 to 800 km/s, are similar (at least statistically). In contrast to the temperature, the peaks in density are much more similar to the observed values, reaching over 20 cm$^{-3}$ at one point (DOY: 111 in the observations and DOY 115 in the simulation). Fifth, the magnetic field magnitude is, on average, of similar amplitude; however, the peak values in the model do not reach those observed. It is also interesting to note that the spacecraft transitions from a region of positive to negative polarity at day 110 in both the model and observations, remaining there until day 123 (model) / 124 (observations). 
Overall, we infer that the variability in the plasma and field results is better captured in CR2002 than in CR2063.

\begin{figure}[ht!]
    \centering
    \includegraphics[width=0.666\textwidth]{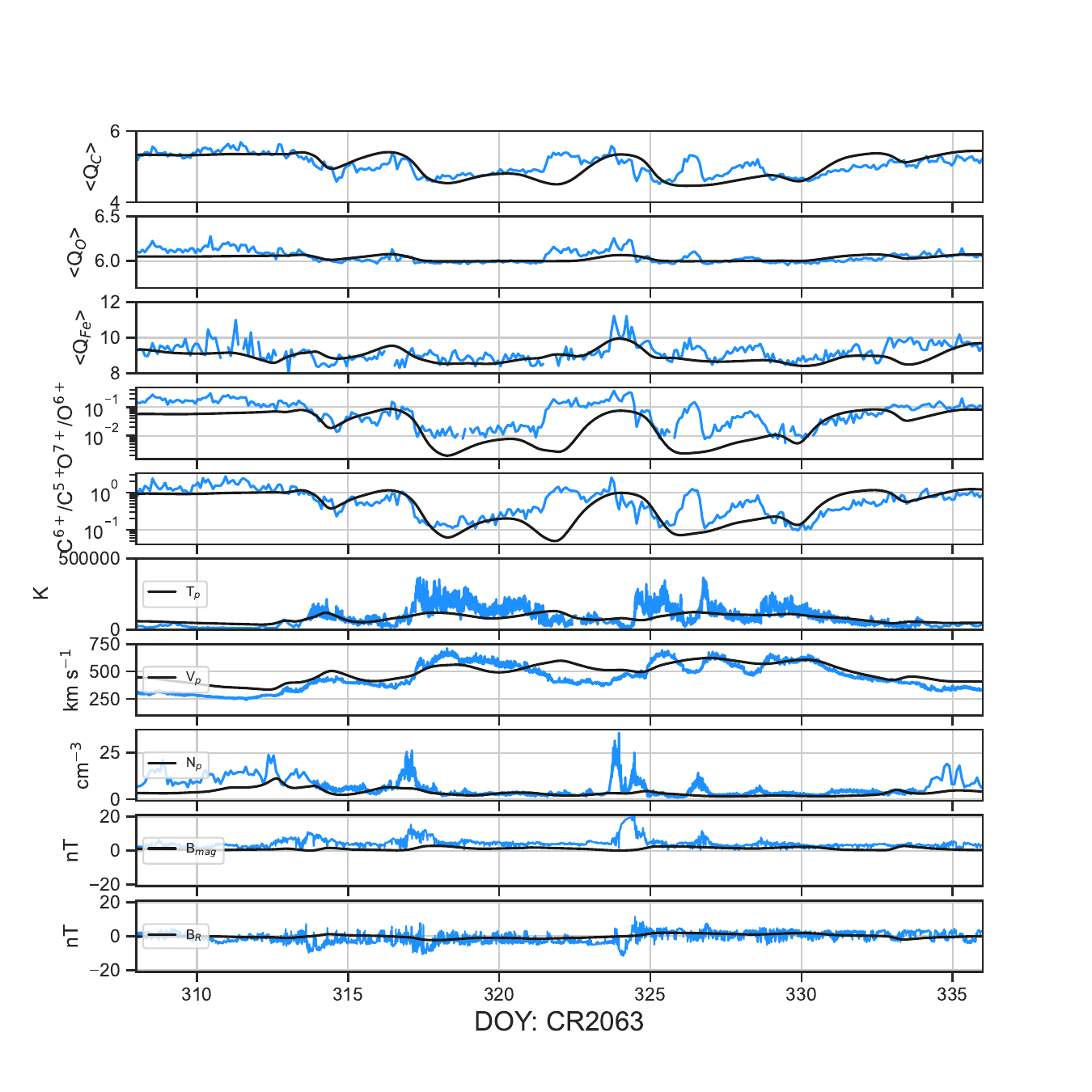}
    \caption{Direct comparison between modeled (black) and observed (blue) in-situ measurements at ACE for CR 2063, corresponding to November 04 to December 02, 2007 (doy: 308 to 336).  
}
    \label{ace-cr2002-obs-simobs-sim-direct-comp-cr2063}
\end{figure}

\begin{figure}[ht!]
    \centering
    \includegraphics[width=0.666\textwidth]{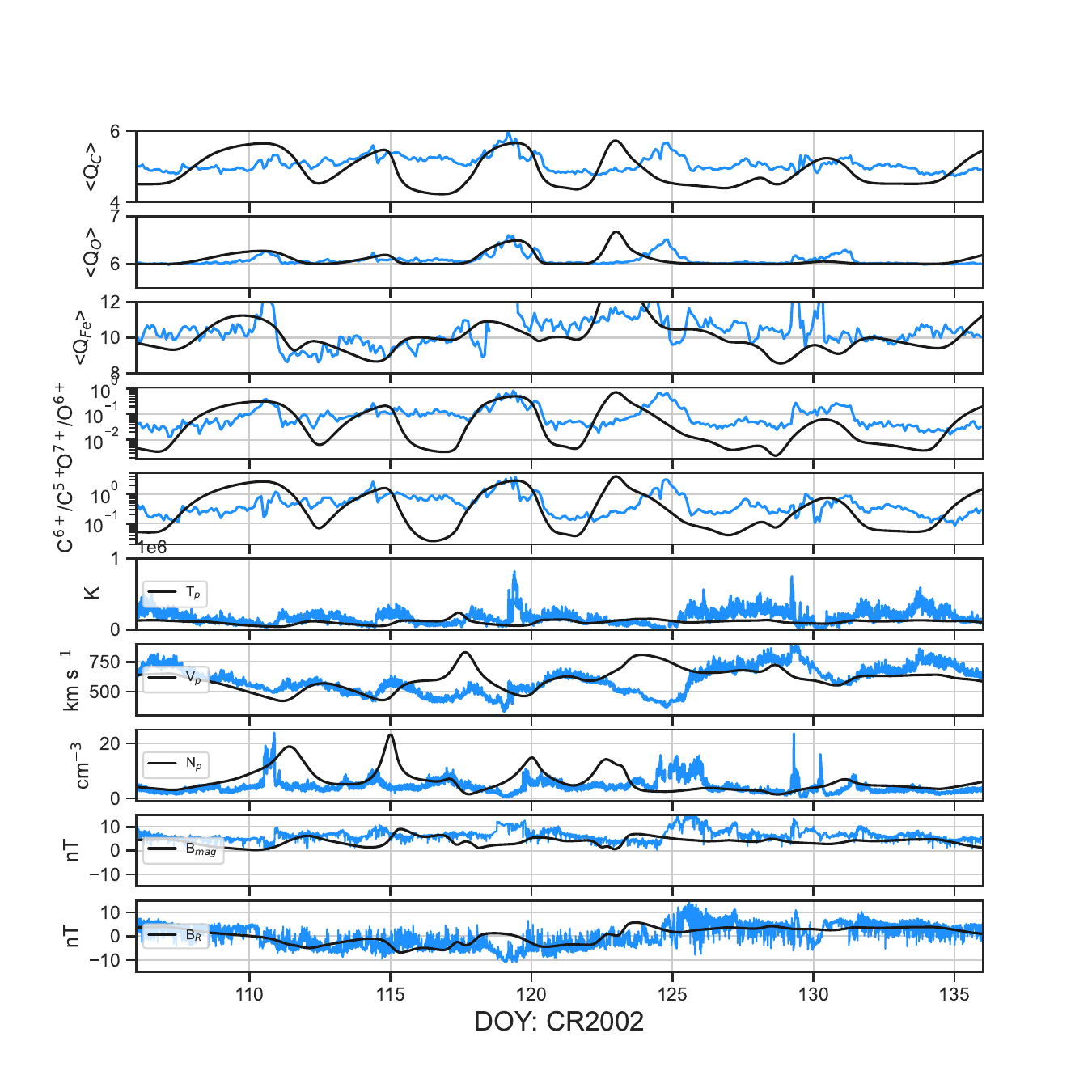}
    \caption{As Figure~\ref{ace-cr2002-obs-simobs-sim-direct-comp-cr2063}, but for CR2002, corresponding to April 15 to May 16, 2002 (doy: 105 to 136).  
}
    \label{ace-cr2002-obs-simobs-sim-direct-comp-cr2002}
\end{figure}

Turning our attention to a more direct comparison, in Figure~\Ref{ace-cr2002-obs-simobs-sim-direct-comp-cr2063} we have overlaid the model time series on top of the measured values for CR2063. While reinforcing some of the points made concerning Figure~\Ref{ace-cr2063-obs-sim}, this comparison also highlights some notable differences between the model and observations. 
First, the average charge state of C, $<Q_C>$ is quite well reproduced by the model. Not only are the absolute values reasonably correct in a statistical sense, but there is some indication that the overall changes as a function of time are captured by the model. This is also true for $<Q_O>$ and $<Q_{Fe}>$ in the second and third panels, although, for the former, the amplitude is smaller than is observed. The ratios, $O^{7+}/O^{6+}$ and $C^{6+}/C^{5+}$, in the next two panels also present a relatively good match with observations. The interval from DOY 315 to 320 and 330 to 335, for example, correspond to (1) low-to-high-speed and (2) declining speed profiles, respectively. Perhaps the biggest disagreement is for the modeled $O^{7+}/O^{6+}$ to underestimate the observed values during the central portion of the interval, where (possibly coincidently) the speed is consistently higher. 
Comparison of the plasma and field components, as noted above, generally shows that the model has not captured the full range of the observations, with the notable exception of the speed profile, which matches well. It is also worth remarking that the density values in the quiescent high-speed wind match the observations well, but the compression regions are not nearly as well developed. This is likely the result of the modeled solar wind streams not showing the equivalent range between low- and high-speed streams. Finally, the lower amplitudes in the modeled field values are a well-known (but, as yet, unresolved) issue with global heliospheric models \citep{linker17a,2019ApJ...884...18R}. 

We now consider the same comparison for CR2002. Here we note several differences, both between the observations and with the solution for CR2063. First, the average charge state of C, $<Q_C>$, varies considerably more in the model than in the observations, with a baseline value below that of the observations. In contrast, the average charge states for O and Fe are much more similar to the measurements both in terms of averages and variations. These features are mirrored in the ratios, $O^{7+}/O^{6+}$ and $C^{6+}/C^{5+}$, where the former is consistently lower than the observations, while the latter tends to match the observations better, at least in terms of average values. It is also worth noting that this is a logarithmic scale. For the plasma comparisons (temperature, speed, density, magnetic field magnitude, and radial component), while there are some overall matches, such as the declining speed profile from DOY 110 to 111/112, and following compression, overall, the model has not captured the dynamical features that were observed. In particular, the low-speed interval centered on DOY 125 is associated with a speed increase in the model. This, in turn, means that the compression that is driven by the higher speed following is not mirrored in the model, which has steady but not faster speed after DOY 125. As we noted previously, the magnetic sector structure is reproduced by the model with the spacecraft being immersed in a negative polarity region from day 111 to 123/125, where the return to positive polarity arrives prematurely in the model. This comparison is consistent with our previous attempts to reproduce the structure of the solar wind near the peak of the solar cycle, a period that is defined by shorter-time-scale evolution of the photospheric magnetic fields, an evolution that cannot be captured by CR-averaged maps. 

Contrasting CR2002 with CR2063, we remark that the former produces better matches with the plasma and field profiles but variations that are too strong in the charge state values. On the other hand, the latter produces much more damped variations in the plasma and field variations but charge state variations that are much more like the observations. Overall, however, the CR2063 comparison appears to have captured the dynamical features of the solar wind during this interval better than CR2002, which is what we would have anticipated given their relative positions in the phase of the solar cycle.   

\begin{figure}[ht!]
    \centering
    \includegraphics[width=0.95\textwidth]{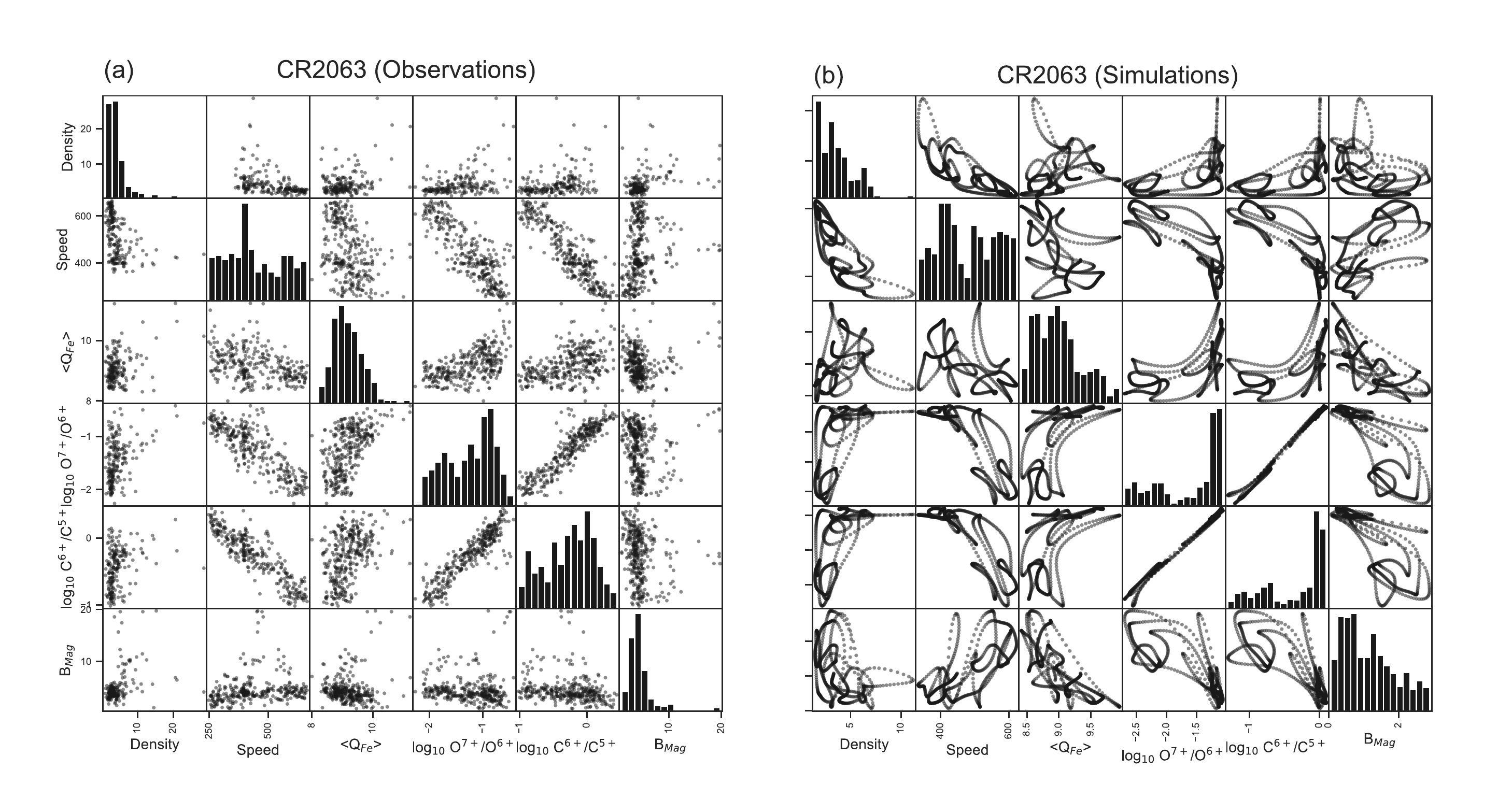}
    \caption{Scatterplot matrices for two-hour averaged (a) observations and (b) model results of density, speed, the average charge state of iron, $O^{7+}/O^{6+}$, $C^{6+}/C^{5+}$, and the magnitude of the magnetic field for CR2063. The diagonal panel summarises the distribution of that variable.   
}
    \label{scatterplot_cr2063}
\end{figure}

\begin{figure}[ht!]
    \centering
    \includegraphics[width=0.95\textwidth]{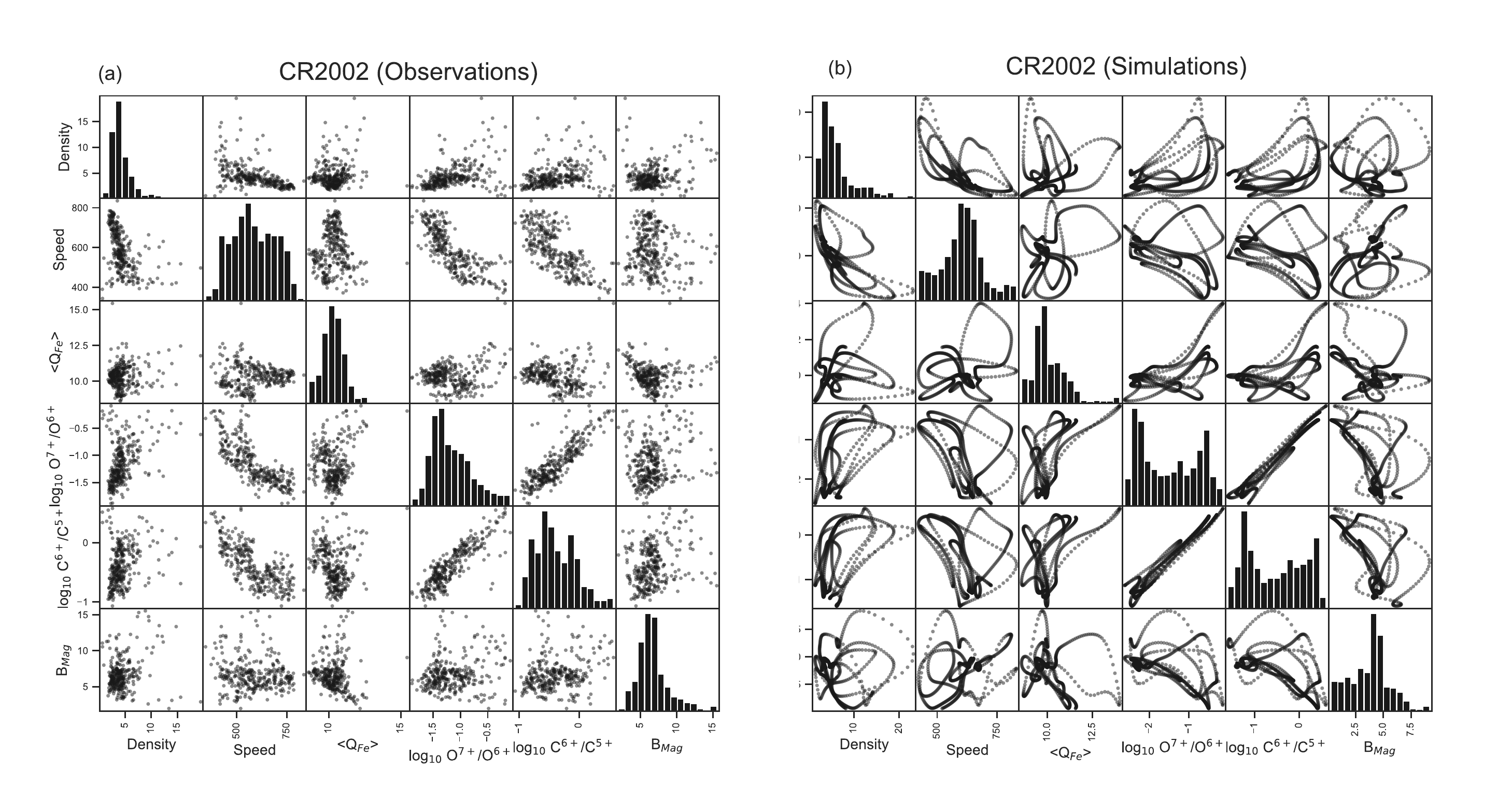}
    \caption{As Figure~\ref{scatterplot_cr2063} but for CR2002.    
}
    \label{scatterplot_cr2002}
\end{figure}

Finally, we consider the statistical properties of the model output and a comparison with observations. By limiting our comparison to these specific CRs, we acknowledge that these are snapshots over a relatively short interval of time. Thus, they are not intended to provide any general statistical properties of the solar wind at a particular phase of the solar cycle but a measure of the (lack of) agreement between the model results and the observations. Figure~\ref{scatterplot_cr2063} shows scatterplot matrices for (a) observations and (b) model results of density, speed, the average charge state of iron, $O^{7+}/O^{6+}$, $C^{6+}/C^{5+}$, and the magnitude of the magnetic field. The diagonal panel summarises the distribution of that variable. Focusing first on this diagonal and using the observations as the ``ground truth'' to validate the model results, we note the following from panel (a). 
First, most parameters show a roughly unimodal distribution (i.e., a single peak), with density, $<Q_{Fe}>$, and $B_{mag}$ showing log-normal (in the sense of being right-skewed) distributions. The speed profile, on the other hand, is relatively flat. The charge-state ratios $O^{7+}/O^{6+}$ and $C^{6+}/C^{5+}$ could be interpreted as being bimodal, although it could also be argued that they are merely broader and noisier distributions. Comparing these with the model histograms, we see that the overall peaks and relative distributions in the modeled parameters are in reasonable agreement with the observations (note that the x-axis ranges have been adjusted to accentuate the variation in each parameter and are not necessarily the same between observation and model parameter).  

Again, we emphasize the strong caveat of drawing too many inferences from this limited comparison. In addition to only sampling a small solid angle of the total solar wind expelled by the sun during this time period, more importantly, the hypothetical spacecraft that fly through the model did not sample the same wind, statistically as was observed. The difference in the speed profiles, for example, provides a reasonable explanation for the mismatch in the charge state ratios. Despite this, it is worth remarking that the overall peaks and relative distributions in the modeled parameters are in reasonable agreement with the observations. The peak average state of Fe, for example, is well-matched. 

Moving on to the off-diagonal panels, we note a strong correlation between $O^{7+}/O^{6+}$ and $C^{6+}/C^{5+}$ in both the observations and simulations. In fact, the latter shows an even stronger correlation, likely due to either noise or processes that the model does not account for, present in the data. There are also albeit less pronounced negative correlations between $O^{7+}/O^{6+}$ / $C^{6+}/C^{5+}$ and speed. $<Q_{Fe}>$, on the other hand, appears only marginally correlated with speed.  

A final point worth making from Figure~\ref{scatterplot_cr2063} (b) is the apparent hysteresis present amongst the variables. There are several possible explanations for this, including simplified model dynamics, artificial feedback mechanisms, data noise, and variability, or non-linear behavior in the model that is less pronounced in reality. A visual scan of some of the panels in (a) suggests that there is some coherency from point to point (e.g., straight-line and arc patterns); thus, it may be that at least some hysteresis is present but masked underlying noise and/or physical processes that are present in the observations but absent from the simulations. 

Next, we consider the statistical properties of the wind during CR2002 in Figure~\ref{scatterplot_cr2002}. 
In this case, all parameters show a roughly unimodal distribution, with density, $B_{mag}$, $O^{7+}/O^{6+}$, and $C^{6+}/C^{5+}$ showing log-normal distributions. The speed profile, on the other hand, is much broader. While the modeled density is also log-normal, the other modeled parameters deviate from the observed distributions: Speed is more peaked, $O^{7+}/O^{6+}$ and $C^{6+}/C^{5+}$ seem to be bimodal, and $B_{mag}$ is relatively flat, except for a sharp peak at 5 nT.  

Moving on to the off-diagonal panels, we again note a relatively strong correlation between $O^{7+}/O^{6+}$ and $C^{6+}/C^{5+}$ in both the observations and simulations. Similarly, there is are weaker negative correlations between $O^{7+}/O^{6+}$ / $C^{6+}/C^{5+}$ and speed. $<Q_{Fe}>$, again, does not appear to be correlated with speed. In this case, however, the data do seem to show two clusters: One for low-speed/low $<Q_{Fe}>$ and another with higher-speed/higher-$<Q_{Fe}>$. Each cluster shows a potentially negative correlation between speed and $<Q_{Fe}>$. Given the limited sample size, however, we caution that this is, at best, a speculative inference. 

\begin{figure}[ht!]
    \centering
    \includegraphics[width=0.95\textwidth]{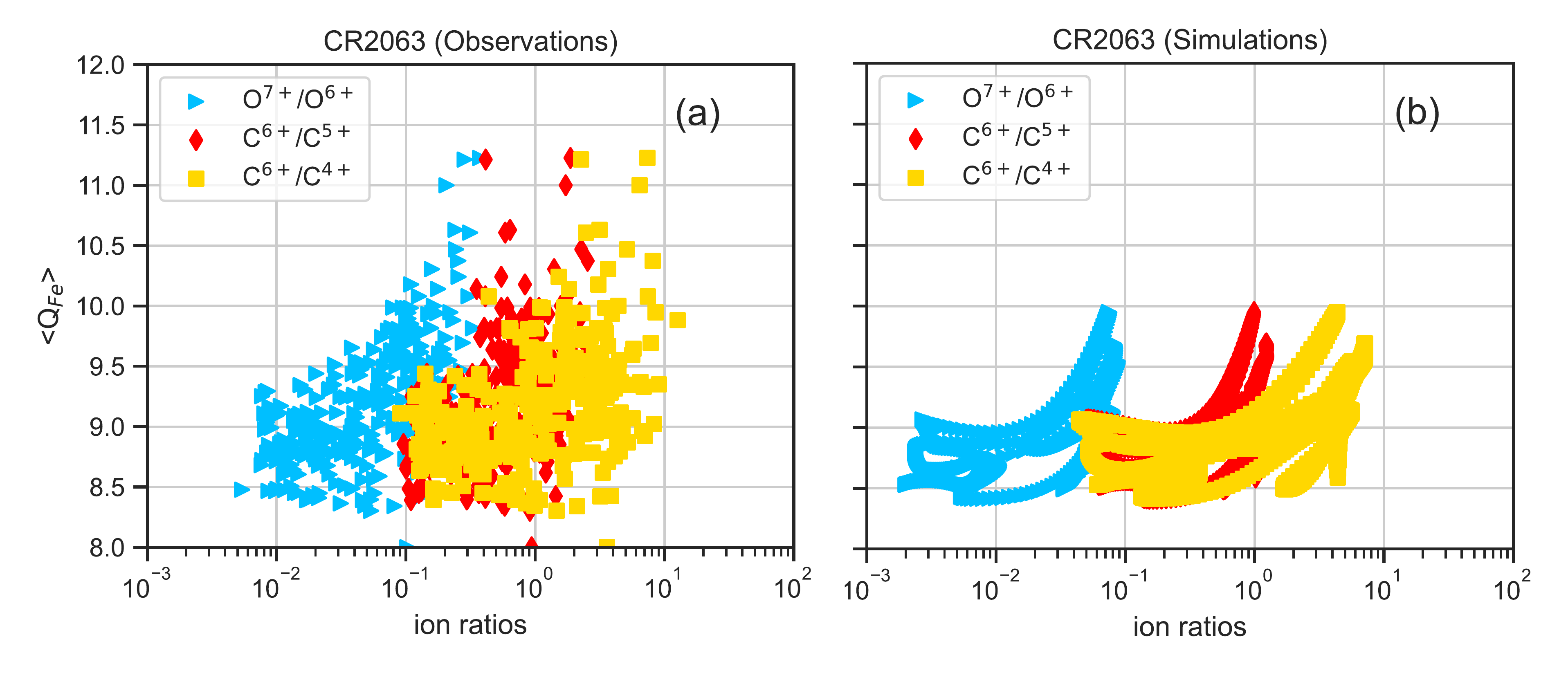}
    \caption{Average charge state of iron ($<Q_{Fe}>$) versus the ion ratios for $O^{7+}/O^{6+}$ (blue), $C^{6+}/C^{5+}$ (red), and $C^{6+}/C^{4+}$ (yellow) from the observations (a) and simulation (b) for CR2063.   
}
    \label{fe-ratios-cr2063}
\end{figure}

As a final comparison between the charge-state datasets, in Figure~\ref{fe-ratios-cr2063}, we compare the ion ratios for $O^{7+}/O^{6+}$ (blue), $C^{6+}/C^{5+}$ (red), and $C^{6+}/C^{4+}$ (yellow) against $<Q_{Fe}>$ for CR2063. Both observations and model show that there is no obvious dependence of $<Q_{Fe}>$ on a particular ratio; however, the centroids of the ratios themselves are somewhat separated, although the  $C^{6+}/C^{5+}$ (red), and $C^{6+}/C^{4+}$ overlap significantly. This pattern is reproduced in the simulation results, and, in fact, the model output even suggests an upward curving of adjacent plasma (a hysteresis effect), which, in retrospect, could be inferred from the data. Again, we note that the model results do not capture the full variability in   $<Q_{Fe}>$ during this interval, showing maximum charge states of 10. On the other hand, the ion ratios show a breadth that is at least as large as the observations. 

\begin{figure}[ht!]
    \centering
    \includegraphics[width=0.95\textwidth]{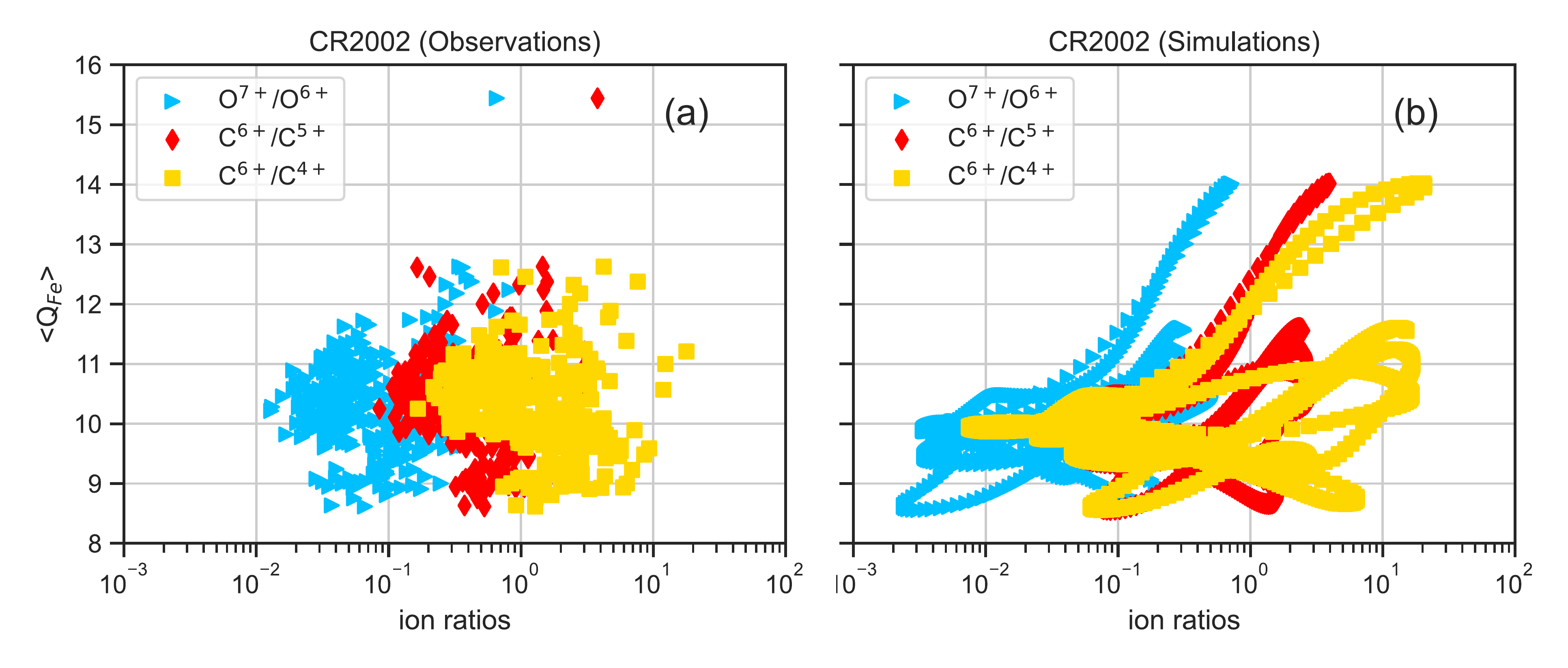}
    \caption{As Figure~\ref{fe-ratios-cr2063}, but for CR2002.   
}
    \label{fe-ratios-cr2002}
\end{figure}

In Figure~\ref{fe-ratios-cr2002} we show the same comparison but for CR2002. Many of the same points can be made. In this case, however, the model results are even broader than the observations would suggest. As with CR2063, the centroids of the observed and modeled distributions match reasonably well. The hysteresis effect, although subdued, remains, suggesting a roughly linear relationship between neighboring parcels of plasma. This is not present in the observed clusters but may be masked by additional processes or uncertainties in the measurements. 

\section{Conclusions and Discussion}

In this study, we have presented a detailed analysis of solar wind charge states by integrating a non-equilibrium ionization model with a global MHD model, for both solar minimum and maximum conditions. Our results demonstrate that the model can accurately capture the key features of the observed charge states, offering a significant improvement over previous 1-D calculations. By resolving the discrepancies noted in earlier models, our approach provides a robust framework for tracing the evolution of charge states from the corona into the solar wind. 

Several potentially important limitations should be considered when interpreting these results and the inferences made. First, the MHD solution used to drive the charge state calculations was in a steady-state equilibrium. As such, it did not include transient phenomena, such as interchange reconnection, which may play a fundamental role in the production, or at least modulation, of the charge state profiles. Intriguingly, however, given the reasonable matches between the observations and model results, particularly for CR2063, this may suggest that such processes may not be defining or that the current model is somehow capturing their effects, at least to the extent of reproducing these in-situ measurements. One way to address this would be to calculate charge state distributions in a fully time-dependent simulation, such as was recently completed for the total solar eclipse of 2024-04-08 \citep{downs24a}. 

Second, our results are no doubt sensitive to the specific intervals we studied. While they may be representative, we are cautious about over-interpreting the results. For example, at solar minimum (CR6063), the variability of charge states is sometimes too low, whereas at solar maximum (CR2002), the amplitudes are too large. Is this a result of these specific rotations or a limitation of the model prescription, such as the parameters used to drive the WTD heating and acceleration? To resolve this requires the study of more intervals; a set that spans the entire solar cycle. 

Third, while the ambient solar wind produced by the model is generally representative of the quality of MHD solutions, these two examples are not the quintessential best cases. Thus, to remove a further source of error, we could identify Carrington rotations for which the MHD solution is an accurate facsimile of the in-situ measurements. Examples such as CR1912/1913 (aka, the ``Whole Sun Month'' WSM \citep{riley01a}), where the plasma and field measurements from even simpler models were shown to be excellent matches with the data, could remove this specific source of uncertainty. Fourth, the MHD model, which has been developed and refined over more than 30 years, in some sense, represents an average model of the solar corona over this period. The heliospheric model, for example, relies upon model parameters that were ``tuned'' on data from the end of cycle 22 and the start of cycle 23. Since then, it appears that the Sun has undergone a secular change in activity (see the evolution of peak amplitudes in Figure~\ref{data-availability}). Thus, some of the mismatch between CR2063 and CR2002 may be due to us not accounting for this in the model. Fifth, the underestimation of plasma parameter variability by the model suggests that dynamic interactions and/or transient phenomena occurring in the solar wind are not fully captured. 

Our work can be compared and contrasted with several other studies. First, the present study is a natural extension of our early 1-D study \citep{lionello19a}. In that work, we implemented a similar non-equilibrium ionization code coupled with a 1D WTD hydrodynamic solar wind model. We note that, while informative, 1-D models are intrinsically limited in that they contain no coronal magnetic field or its associated 3-D structure. We found that when compared with Ulysses observations, the charge state ratios from the 1-D solutions were too low. We proposed that the modeled flow speed in the low corona was too high.  Because of this, the ions propagated through the region too quickly to evolve the observed ratios. By heuristically reducing this speed, we were able to bring the model results into agreement with the observations. That disagreement has largely disappeared in the current 3-D results. There are several possible reasons for this. For example, it is possible that the current investigation, which focuses on in-ecliptic measurements, may not be as sensitive to this. There are, however, intervals of high-speed wind during both CR 2063 and 2002 that do not manifest this deficit; however, there are also intervals of disagreement. Assuming that the issue has been resolved, or at least mitigated, higher charge state ratios can be produced in the model by (1) stronger photospheric magnetic fields, which heat the plasma more, hence raising its temperature, (2) higher densities, which effectively increase the scale height at which the ions ``freeze-in'', or (3) lower plasma speeds, which increase the time over which the ions can ionize. To fully resolve would require a more comprehensive set of simulations to compare with Ulysses observations, as discussed below. 

The research team at the University of Michigan has also developed an increasingly sophisticated MHD-charge-state modeling suite. In their transition from 1-D to 3-D \citep{landi14a,oran15a,szente22a}, they also found a deficit between the modeled and observed charge state ratios. Based on earlier suggestions by \citet{esser00a,esser01a,esser02a}, they modeled the effects of suprathermal tails on the electron distributions on the resulting charge state ratios. They found that including this contribution substantially resolved the mismatch. As a side note, \citet{esser00a,esser01a,esser02a} also proposed ion differential flows as a source for the discrepancy (which in their case was between coronal temperatures derived from in-situ charge states and those inferred from spectroscopic (coronal) images). This remains to be tested by global MHD/charge state models. 

There are several potentially fruitful directions to pursue based on this initial study. For example, while we focused on interpreting ACE/SWICS results during two specific Carrington rotations, the model comparison could be extended to include Ulysses measurements as well. This would provide a broader context for interpreting the model results; however, it would require additional simulations to be completed, particularly during the slower motion towards and away from aphelion. Nevertheless, such a comparison could give strong additional constraints from vastly different latitudinal regions during these longer periods. Similarly, if the caveats associated with the v2.0 SWICS data can be reasonably addressed, comparisons with Solar Orbiter and ACE charge state measurements could provide longitudinal constraints on the overall model solutions (complementary to previous studies using plasma and magnetic field in situ measurements). 

The new time-dependent simulations of the April 8th, 2024 eclipse \citep{downs24a} open up additional possibilities of exploring how charge-state distributions could be formed directly from the opening and/or closing of magnetic field lines in the corona. Unlike the steady-state solution approach, for which the charge state calculations can be completed after the MHD run is completed, for the time-dependent case, the charge state module must be run in tandem. A second obvious avenue to pursue is to build up a larger repository of model and observation comparisons. It should cover a wide range of phases of the solar cycle and also track the secular evolution of the Sun from the early 1990's through the Solar Orbiter interval. This will allow us to identify any fundamental limitations in the modeling as well as any biases in the model output. Finally, and more pragmatically, during the interval from 2012 (the date of the start of SWICSv2.0) through the present, the ACE/SWICS dataset has some known saturation issues that have limited its usefulness. To the extent that these data can be used to calibrate and constrain the global model output, this would add a valuable set of constraints during most of cycle 24 and cycle 25.   

This study is a first step in understanding the observed properties of charge states in the solar wind and relating them to global model results. Ultimately, we believe these types of comparisons will provide crucial feedback for improving the localized heating and acceleration processes in the model. When combined with other observational constraints, such as EUV and white light images, and plasma and magnetic field in-situ measurements, these charge-state datasets will not only improve our scientific understanding but, ultimately, lead to more accurate operational space weather forecasts.

\begin{acknowledgments}
PR gratefully acknowledges support from NASA (80NSSC18K0100, NNX16AG86G, 80NSSC18K1129, 80NSSC18K0101, 80NSSC20K1285, 80NSSC18K1201, and NNN06AA01C), NOAA (NA18NWS4680081), and the US Air Force (FA9550-15-C-0001). RL acknowledges support from NASA (80NSSC20K0192).
\end{acknowledgments}

%



\software{MAS \citep{2018NatAs...2..913M}, 
          Charge-State-Code \citep{2015A&C....12....1S,shen24a},
          Python packages \citep{10.5555/1593511}, gnuplot \citep{{gnuplot}}.
          }






\bibliographystyle{aasjournal}





\end{document}